\documentclass[aps,prb,preprint,showpacs,preprintnumbers,amsmath,amssymb]{revtex4}
\usepackage{graphicx}
\usepackage{color}
\usepackage{float}
\begin{document}
\draft
\title{Many-body theory for proton-induced point-defect effects on losses of electron energy and photons in quantum wells}

\author{Danhong Huang$^{1}$, Andrii Iurov$^{2}$, Fei Gao$^{3}$, Godfrey Gumbs$^{4}$ and D. A. Cardimona$^{1}$}
\affiliation{
$^{1}$Air Force Research Laboratory, Space Vehicles Directorate, Kirtland Air Force Base, NM 87117, USA\\
$^{2}$Center for High Technology Materials, University of New Mexico, 1313 Goddard SE, Albuquerque, NM, 87106, USA\\
$^{3}$Department of Nuclear Engineering and Radiological Sciences, University of Michigan, 500 S. State Street, Ann Arbor, MI 48109, USA\\
$^{4}$Department of Physics and Astronomy, Hunter College of the City University of New York, 695 Park Avenue, New York, NY 10065, USA}

\date{\today}

\begin{abstract}
The effects of point defects on the loss of either energies of ballistic electron beams or incident photons are studied by using a many-body theory in a multi-quantum-well system. This includes
the defect-induced vertex correction to a bare polarization function of electrons within the ladder approximation as well as
the intralayer and interlayer screening of defect-electron interactions are also taken into account in the random-phase approximation.
The numerical results of defect effects on both energy-loss and optical-absorption spectra are presented and analyzed for various defect densities, number of quantum wells, and wave vectors.  
The diffusion-reaction equation is employed for calculating distributions of point defects in a layered structure. For completeness,
the production rate for Frenkel-pair defects and their initial concentration are obtained based on atomic-level molecular-dynamics simulations.
By combining defect-effect, diffusion-reaction and molecular-dynamics models proposed in this paper with a space-weather forecast model for the first time,
it will be possible to enable specific designing for electronic and optoelectronic quantum devices that will be operated in space with radiation-hardening protection, and therefore, will
effectively extend the lifetime of these satellite onboard electronic and optoelectronic devices. 	
\end{abstract}
\pacs{}
\maketitle

\section{Introduction}
\label{sec-1}

Point defects (vacancies and interstitials) are produced by displacements of atoms from their thermal-equilibrium lattice sites,\,\cite{book-gary,book-sigmund}
where the lattice-atom displacements are mainly caused by a proton-irradiation induced
primary knock-on atom (PKA) on a time scale shorter than $100$\,ps for building up point defects without thermal reactions.
These initial displacements are followed immediately by defect mutual recombinations or reactions with sinks
(clustering or dissolution of clusters for point-defect stabilizations)\,\cite{add2,gao6} on a time scale shorter than $10\,$ns,
then possibly by thermally-activated defect migrations\,\cite{gao5} up to a time scale much longer than $10$\,ns (steady-state distributions).
Such atom displacements depend not only on the energy-dependent flux of protons but also on the differential energy transfer cross sections (probabilities) for collision between
atoms, interatomic Coulomb interactions and even kinetic-energy loss to core-level electrons of an atom (ionizations). The sample temperature at which the irradiation has been done
also significantly affects the diffusion of defects,
their stability as clusters and the formation of Frenkel pairs.\,\cite{gao7}
One of the effective calculation methods for studying the non-thermal spatial-temporal distributions of proton-irradiation-induced point defects is the molecular-dynamics (MD) model\,\cite{gao2}. However,
the system size increases quadratically with the initial kinetic energy of protons and the time scale can easily run up to several hundred
picoseconds. In this case, the defect reaction process driven by thermal migration cannot be included in the MD model due to its much longer time scale. Practically,
if the system time evolution goes above $100$\,ps, either the kinetic lattice Monte-Carlo\,\cite{gao1} or the diffusion-reaction equation\,\cite{ryazanov,stoller} method should be used instead.
\medskip

In the presence of defects, dangling bonds attached to these point defects can capture Bloch electrons through multi-phonon emission to form localized charged centers.
The randomly-distributed charge centers will further affect electron responses to either an external ballistic electron beam\,\cite{book} or incident photons\,\cite{add-1}.
Physically, the defect modifications to the electron response function can be addressed by a vertex correction\,\cite{add-1} to a bare electron polarization function in the ladder approximation\,\cite{stern} (LA).
In addition, both the intralayer and interlayer screening corrections in a multi-quantum-well system can be included by using the random-phase approximation\,\cite{stern,add-3} (RPA).
The many-body theory presented here is crucial for understanding the full mechanism for characterizing defects,\,\cite{hallen} defect effects,\,\cite{doan}
as well as for developing effective mitigation in early design stages of electronic devices. Equipped with this multi-timescale microscopic theory,\,\cite{ajss} the experimental characterization of post-irradiated
test devices\,\cite{hubbs} is able to provide useful information on the device architecture's susceptibility to space radiation effects\,\cite{simoen}. Furthermore, our physics model should also allow for accurate
prediction of device-performance degradation by using the
space weather forecast\,\cite{swf,cooke} for a particular orbit. With this paper, we expect to bridge the gap between researchers studying radiation-induced damage in materials\,\cite{book-gary,book-sigmund,add-4,add-5}
and others characterizing irradiation-induced
performance degradation in devices.\,\cite{vince,chris}
\medskip

The rest of the paper is organized as follows. In Sec.\ \ref{sec-2}, we present our theoretical model and numerical results
to highlight the defect effects on losses of electron energy and photons in multi-quantum-well systems, 
where defect potentials and vertex corrections, defect effects on partial and total polarization functions, electron-energy loss functions and intrasubband and intersubband absorption spectra have
been demonstrated and analyzed.
In Sec.\ \ref{sec-3}, ultrafast dynamics related to defect production, as well as the follow-up defect diffusion and reaction, will be studied and a steady-state one-dimensional
distribution function of point defects will be calculated to provide a direct input for modeling defect effects discussed in Sec.\,\ref{sec-2}.
Finally, a summary and some remarks are presented in Sec.\ \ref{sec-4}.

\section{Effects of point defects}
\label{sec-2}

In Sect.\,\ref{sec-2}, we first look into effects of point defects on the electron polarization function in a single wide quantum well. After generalizing the system to multiple quantum wells,
we further study the kinetic-energy loss of a parallel (or perpendicular) electron beam. For a comparison, we also calculate the loss of incident photons
with a field polarization parallel (or perpendicular) to the quantum-well planes, corresponding to intrasubband\,\cite{add-13} (or intersubband\,\cite{add-14}) optical transitions of electrons, respectively.

\subsection{Effects on Electron Polarization Function}
\label{sec-2-1}

Since the wave functions of individual point defects are spatially localized, we expect that the interaction between electrons and charged point defects can only affect the screening to the intralayer Coulomb interaction.
Therefore, we start with a study of defect effects in a single quantum well. The exchange-interaction-induced vertex correction to a bare polarization function of electrons in a quantum well has been addressed before\,\cite{add-1} 
within the ladder approximation.
\medskip

For an $n$-doped quantum well, the total electron polarization function\,\cite{add-11} can be written as a sum of partial polarization functions,
i.e., $\displaystyle{\tilde{\chi}(q_\|,\,\omega)=\sum_{n\leq n'}\,\chi_{n,n'}(q_\|,\,\omega)}$,
where $q_\|$ is an electron wavenumber, $\omega$ is an angular frequency of an electrical (or optical) perturbation, and $n\leq n'=1,\,2,\,\cdots$ label different energy subbands.
Here, each partial polarization function $\chi_{n,n'}(q_\|,\,\omega)$ can be calculated through an inverse dielectric function ${\cal K}_{n,n';\,m,m'}(q_\|,\,\omega)$, according to\,\cite{book}

\begin{equation}
\chi_{n,n'}(q_\|,\,\omega)=\sum_{m\leq m'}\,{\cal K}_{n,n';\,m,m'}(q_\|,\,\omega)\,\chi^{(0)}_{m,m'}(q_\|,\,\omega)\,
\Gamma_{m,m'}(q_\|,\,\omega)\ ,
\label{eqn-1}
\end{equation}
where the second term is a defect correction, and the bare polarization function $\chi^{(0)}_{n,n'}(q_\|,\,\omega)$ takes the form

\[
\chi^{(0)}_{m,m'}(q_\|,\,\omega)=\frac{1}{2\pi^2}\int\limits_0^\infty k_\|dk_\|\int\limits_0^{2\pi} d\theta_{{\bf k}_\|,{\bf q}_\|}
\]
\begin{equation}
\times\left\{\frac{f_0[\varepsilon_{m}(k_\|)]-f_0[\varepsilon_{m'}(|{\bf k}_\|+{\bf q}_\||)]}{\hbar\omega+i\gamma_0-\varepsilon_{m'}(|{\bf k}_\|+{\bf q}_\||)+\varepsilon_{m}(k_\|)}
+\frac{f_0[\varepsilon_{m'}(|{\bf k}_\|+{\bf q}_\||)]-f_0[\varepsilon_{m}(k_\|)]}{\hbar\omega+i\gamma_0-\varepsilon_{m}(k_\|)+\varepsilon_{m'}(|{\bf k}_\|+{\bf q}_\||)}\right\}\ ,
\label{eqn-2}
\end{equation}
$\theta_{{\bf k}_\|,{\bf q}_\|}$ is the angle between wave vectors ${\bf k}_\|$ and ${\bf q}_\|$,
$\gamma_0$ is the level broadening, $\varepsilon_{n}(k_\|)=\varepsilon_n+\hbar^2 k_\|^2/2\mu^\ast$ are subband energies, $\varepsilon_n=\pi^2\hbar^2n^2/2\mu^\ast L_W^2$,
$\mu^\ast$ is the  effective mass, $L_W$ is the well width,
$f_0(x)=\{1+\exp[(x-u_{c})/k_BT]\}^{-1}$ is the Fermi function, $u_{c}$ and $T$ are the chemical potential and temperature of electrons, respectively.
\medskip

\begin{figure}
\includegraphics[width=0.75\textwidth]{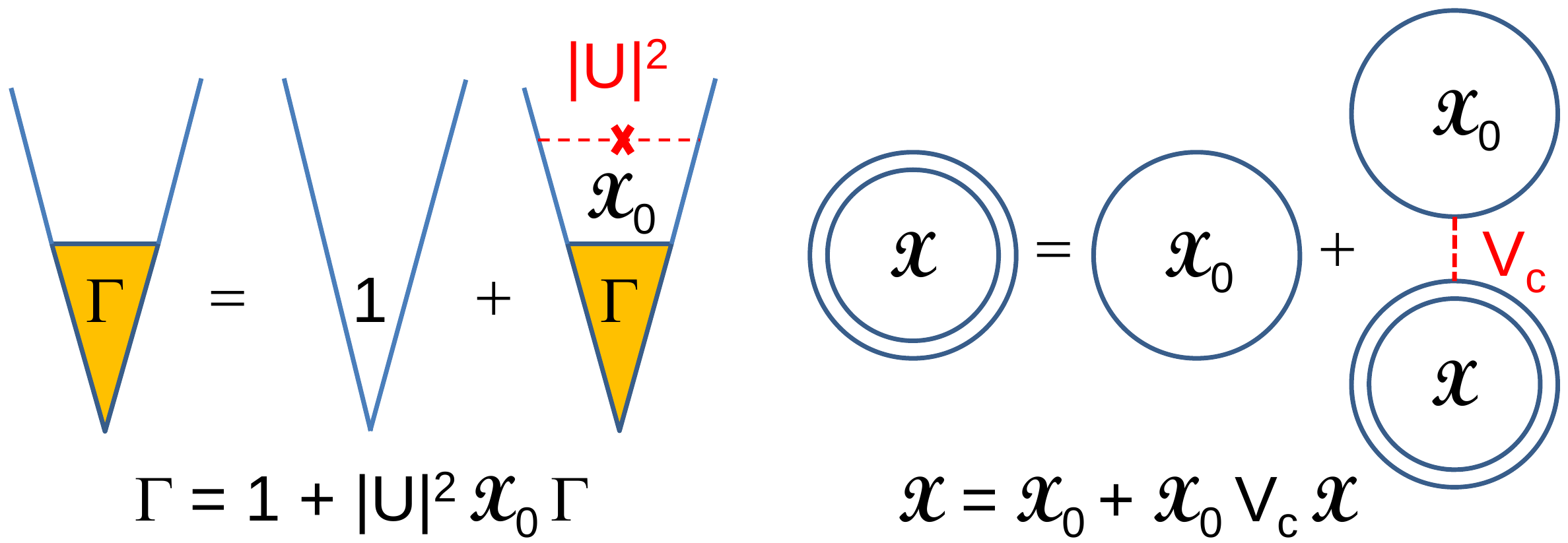}
\caption{(Color online) (Left) Graphic representation for the ladder approximation used in Eq.\,(\ref{eqn-7});
(Right) graphic representation for the random-phase approximation employed in Eq.\,(\ref{eqn-20}).}
\label{approx}
\end{figure}

In addition, the inverse dielectric function ${\cal K}_{\ell,\ell';\,m,m'}(q_\|,\,\omega)$ in Eq.\,(\ref{eqn-1}) satisfies

\begin{equation}
\sum_{m\leq m'}\,{\cal K}_{\ell,\ell';\,m,m'}(q_\|,\,\omega)\,\epsilon_{m,m';\,n,n'}(q_\|,\,\omega)=\delta_{\ell,n}\,\delta_{\ell',n'}\ ,
\label{eqn-3}
\end{equation}
where $\epsilon_{m,m';\,n,n'}(q_\|,\,\omega)$ is the dielectric function and can be calculated
within the RPA\,\cite{add-3} as (right panel of Fig.\,\ref{approx})

\begin{equation}
\epsilon_{m,m';\,n,n'}(q_\|,\,\omega)=\delta_{m,n}\,\delta_{m',n'}
-\chi^{(0)}_{n,n'}(q_\|,\,\omega)\,\Gamma_{n,n'}(q_\|,\,\omega)\,{\cal V}_{m,m';\,nn'}(q_\|)\ ,
\label{eqn-4}
\end{equation}
and the second term corresponds to the defect correction.
In Eq.\,(\ref{eqn-4}), ${\cal V}_{m,m';\,n,n'}(q_\|)$ are the intralayer Coulomb matrix elements, given by\,\cite{add-12}

\begin{equation}
{\cal V}_{m,m';\,n,n'}(q_\|)=\frac{e^2}{2\epsilon_0\epsilon_d(q_\|+q_s)}\,\int dz
\int dz'\,\left[{\cal F}_m(z)\right]^\ast{\cal F}_{m'}(z)\,
{\rm e}^{-q_\||z-z'|}\,\left[{\cal F}_n(z')\right]^\ast{\cal F}_{n'}(z')\ ,
\label{eqn-5}
\end{equation}
where $\epsilon_d$ is the host-material dielectric constant, ${\cal F}_n(z)=\sqrt{2/L_W}\,\sin[(n\pi/L_W)(z+L_W/2)]$
is the wave function of the $n$th subband, and

\begin{equation}
q_s=\frac{e^2}{2\pi\epsilon_0\epsilon_d}\,\sum_n\,\int\limits_0^\infty k_\|dk_\|\,
\left(-\frac{\partial f_0[\varepsilon_{n}(k_\|)]}{\partial\varepsilon_{n}(k_\|)}\right)\ ,
\label{eqn-6}
\end{equation}
which plays the role of the inverse of a static screening length.\,\cite{add-12}
\medskip

For the defect-vertex correction\,\cite{add-1} $\Gamma_{n,n'}(q_\|,\,\omega)$ introduced in Eqs.\,(\ref{eqn-1}) and (\ref{eqn-4}),
we find the following self-consistent equation within the LA (left panel of Fig.\,\ref{approx})

\[
\Gamma_{n,n'}(q_\|,\,\omega)=1+\left(\frac{Z^\ast e^2}{2\epsilon_0\epsilon_d}\right)^2\frac{1}{2\pi^2}\int\limits_0^\infty p_\|dp_\|\,\chi^{(0)}_{n,n'}(p_\|,\,\omega)\,\Gamma_{n,n'}(p_\|,\,\omega)\,
\]
\[
\times\delta\left[\varepsilon_{n'}\left(\frac{q_\|}{2}\right)-\varepsilon_{n}\left(\frac{p_\|}{2}\right)\right]\int\limits_{-{\cal L}_0/2}^{{\cal L}_0/2} dz_0\,\rho_d(z_0)\,
\left|U_{n,n'}(q_\|,\,p_\|\vert z_0)\right|^2
\]
\begin{equation}
=1+\left(\frac{Z^\ast e^2}{2\epsilon_0\epsilon_d}\right)^2\frac{2\mu^*}{\pi^2\hbar^2}\,\chi^{(0)}_{n,n'}(q^*_\|,\,\omega)\,\Gamma_{n,n'}(q^*_\|,\,\omega)\,
\left(\int\limits_{-{\cal L}_0/2}^{{\cal L}_0/2} dz_0\,\rho_d(z_0)\,\left|\overline{U}_{n,n'}(q_\|,z_0)\right|^2\right)\ ,
\label{eqn-7}
\end{equation}
where $q^*_\|=\sqrt{q^2_\|+8\mu^*\varepsilon_{n'n}/\hbar^2}$, $\varepsilon_{n'n}=\varepsilon_{n'}-\varepsilon_n\geq 0$, and the defect interaction with electrons
$\left|\overline{U}_{n,n'}(q_\|,\,z_0)\right|^2\equiv\left|U_{n,n'}(q_\|,\,q^*_\|\vert z_0)\right|^2$ is calculated as

\[
\left|\overline{U}_{n,n'}(q_\|,\,z_0)\right|^2
=\int\limits_0^{\pi} d\theta\,\left(\frac{{\rm e}^{-\Delta_{n'n}^2(q_\|,\,\theta)\Lambda_\|^2/4}}{\Delta_{n'n}(q_\|,\,\theta)+q_s}\right)^2
\]
\begin{equation}
\times\left(\int\limits_{0}^{L_W/2} dz\,{\cal F}_{n}(z){\cal F}_{n'}(z)\left[{\rm e}^{-\Delta_{n'n}(q_\|,\,\theta)|z-z_0|}\pm{\rm e}^{-\Delta_{n'n}(q_\|,\,\theta)|z+z_0|}\right]\right)^2\ ,
\label{eqn-8}
\end{equation}
the sign $+$ ($-$) corresponds to the case with $n=n'=1$ or $2$ ($n'=2$ and $n=1$),
${\cal L}_0$ is the system size,
$Z^\ast$ is the trapped charge number of a point defect,
$2\Delta_{n'n}^2(q_\|,\,\theta)=q_\|^2-q_\|\sqrt{q^2_\|+8\mu^*\varepsilon_{n'n}/\hbar^2}\,\cos\theta+4\mu^*\varepsilon_{n'n}/\hbar^2$,
$\Lambda_\|$ is the correlation length for randomly-distributed point defects,
and $\rho_d(z_0)$ stands for the one-dimensional distribution function of point defects to be determined later in Sec.\,\ref{sec-3-1}.
Here, $\displaystyle{\int_{-{\cal L}_0/2}^{{\cal L}_0/2} dz_0\,\rho_d(z_0)\left|\overline{U}_{1,2}(q_\|,\,z_0)\right|^2=0}$ if $\rho_d(z_0)=\rho_d(-z_0)$.
\medskip

The lowest-order approximate result of Eq.\,(\ref{eqn-7}) can be obtained simply by replacing $\Gamma_{n,n'}(p_\|,\,\omega)$ with $1$ on the right-hand side of this equation. Therefore, the correction to
$\Gamma_{n,n'}(q_\|,\,\omega)\approx 1$ becomes proportional to the total number of point defects or integral of $|\overline{U}_{n,n'}|^2$ with respect to $z_0$.
In general, the solution of Eq.\,(\ref{eqn-7}) includes all the higher orders of $|\overline{U}_{n,n'}|^2$ by going beyond the second-order Born approximation\,\cite{huang}.
\medskip

\begin{figure}
\includegraphics[width=0.75\textwidth]{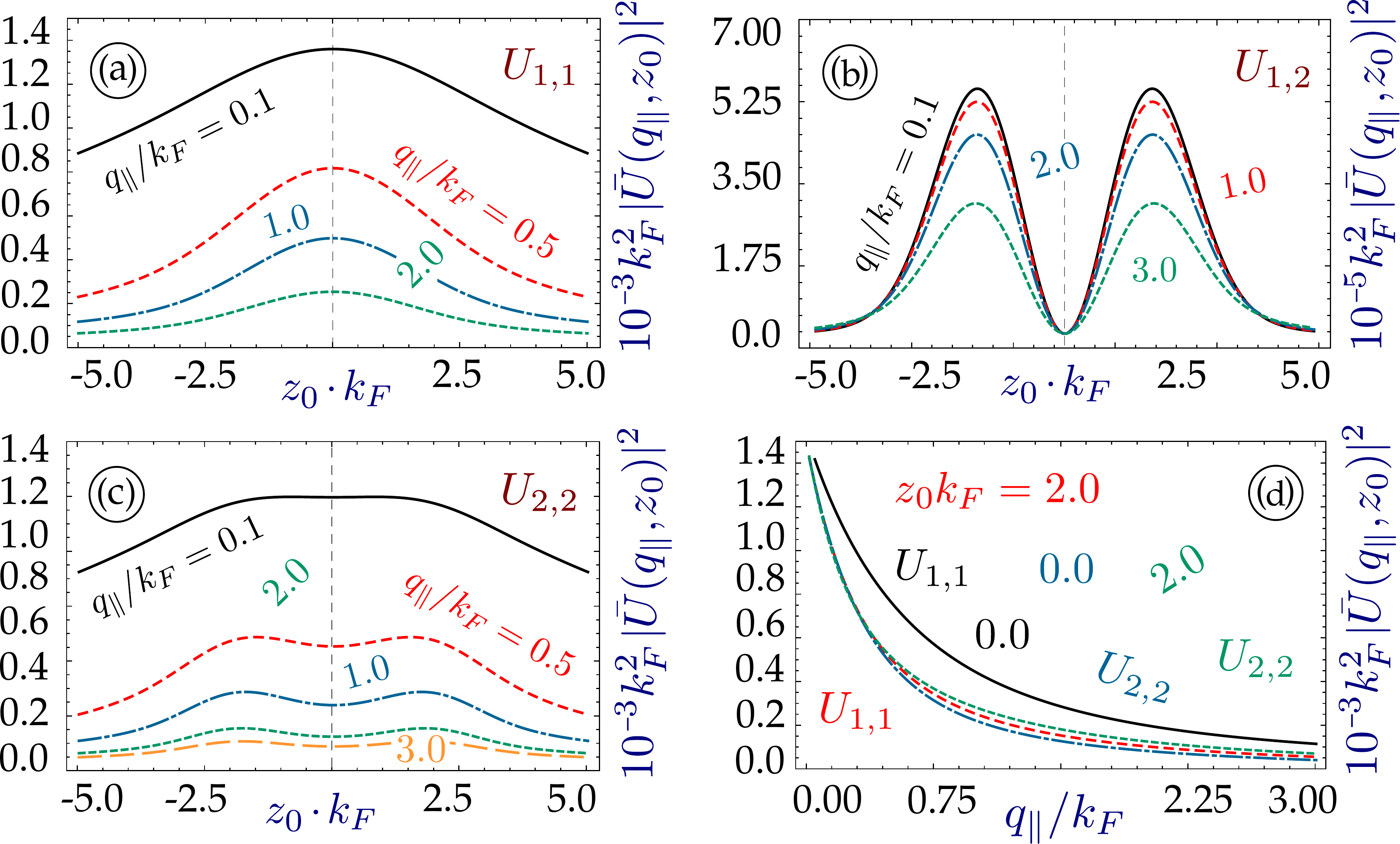}
\caption{(Color online) Defect-electron interaction $|\overline{U}_{n,n'}(q_\|,z_0)|^2$ as a function of point-defect position $z_0$ in ($a$), ($c$)
for $q_\|/k_F=0.1,\,0.5,\,1.0,\,2.0$ and in ($b$) for $q_\|/k_F=0.1,\,1.0,\,2.0,\,3.0$,
as well as a function of electron wave number $q_\|$ in ($d$) for $k_Fz_0=0.0$ and $2.0$. Here,
$k_F=\sqrt{2\pi n_{QW}}$ is the Fermi wave vector, $n_{QW}=1.0\times 10^{11}\,$cm$^{-2}$ is the quantum-well doping density,
$L_W=100\,$nm, $\mu^*=0.067\,m_0$ with free-electron mass $m_0$, $\Lambda_\|=10\,$\AA.}
\label{polar-1}
\end{figure}

The results calculated from Eq.\,(\ref{eqn-8}) for $|\overline{U}_{n,n'}(q_\|,\,z_0)|^2$ are shown in Fig.\,\ref{polar-1},
where the features in $|\overline{U}_{n,n}|^2$ with $n=1,\,2$ for intrasubband interactions
in Figs.\,\ref{polar-1}($a$) and \ref{polar-1}($c$) result from the symmetry
and anti-symmetry properties of the first two electron wave functions in a quantum well.
On the other hand, $|\overline{U}_{1,2}|^2$ in Fig.\,\ref{polar-1}($b$)
for intersubband interactions displays the overlap of these two electron wave functions with opposite symmetries, leading to
two peaks and one node around $z_0=0$. From Figs.\,\ref{polar-1}($a$) and \ref{polar-1}($c$) we further find that both the peak strength and peak width
decrease with increasing $q_\|$, and the reduction of peak strength with $q_\|$ can be seen more clearly from Fig.\,\ref{polar-1}($d$).
In addition, a finite value of $\Delta_{21}(q_\|,\,\theta)$ at $q_\|=0$ leads to a negligible $|\overline{U}_{1,2}|^2$,
and furthermore, the widths of the dual peaks in Fig.\,\ref{polar-1}($b$) spread out significantly with $q_\|$.
\medskip

\begin{figure}
\includegraphics[width=0.75\textwidth]{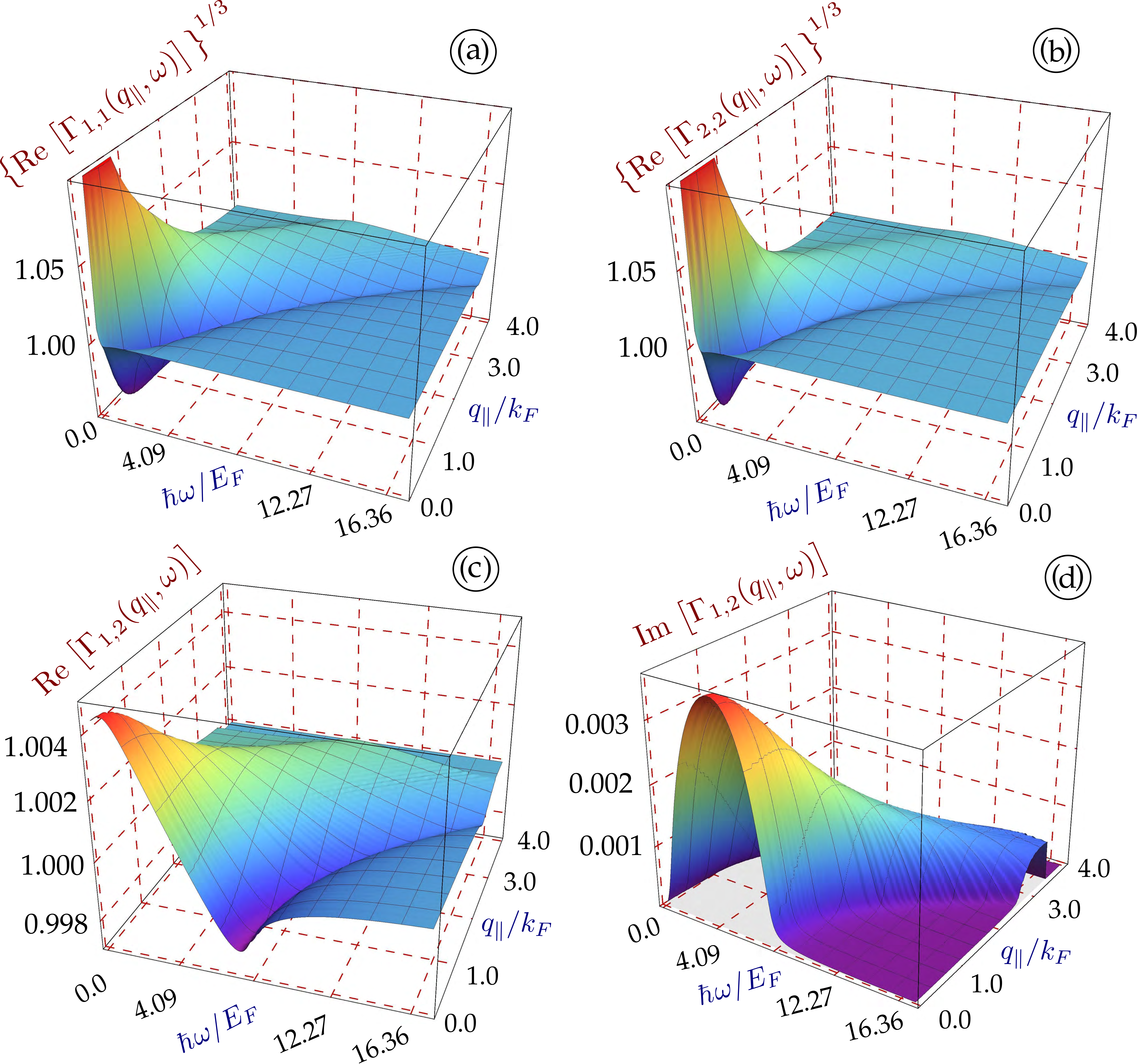}
\caption{(Color online) 3D plots for dynamical defect-vertex correction $\Gamma_{n,n'}(q_\|,\,\omega)$ from the self-consistent solution of Eq.\,(\ref{eqn-7}). Here,
$Z^\ast=1$, $T=4\,$K, $E_F=\hbar^2k_F^2/2\mu^*$, $\epsilon_d=13.3$, ${\cal L}_0/L_W=10$,
$\rho_1=3.0\times 10^{6}\,$cm$^{-1}$, $\rho_2=2.5\times 10^{6}\,$cm$^{-1}$, $\rho_0=1.5\times 10^{6}\,$cm$^{-1}$, $\Delta\rho=1.0\times 10^{6}\,$cm$^{-1}$, and $\kappa=10$. 
The other parameters are the same as those in Fig.\,\ref{polar-1}. 
Results for the real part of $\Gamma_{n,n'}(q_\|,\,\omega)$ with $n=n'=1$, $n=n'=2$ and $n=1,\,n'=2$ are presented in ($a$), ($b$) and ($c$), respectively, while the result for the imaginary part of 
$\Gamma_{1,2}(q_\|,\,\omega)$ is displayed in ($d$). Here, both subbands are occupied.}
\label{polar-2}
\end{figure}

Based on the calculated $|\overline{U}_{n,n'}(q_\|,\,z_0)|^2$ in Fig.\,\ref{polar-1}, Eq.\,(\ref{eqn-7}) can be applied to compute the dynamical defect-vertex correction $\Gamma_{n,n'}(q_\|,\,\omega)$ with respect to unity 
in the ladder approximation. In order to simulate the physical distribution of defects shown in Fig.\,\ref{distri-1}, we assume a regional form, i.e.,
$\rho_d(z_0)/\kappa=\rho_1\Theta(-z_0-L_W/2)]+\rho_2\Theta(z_0-L_W/2)+[\rho_0+z_0(\Delta\rho/L_W)]\Theta(L_W/2-|z_0|)$, 
where $\Theta(x)$ is a unit-step function and $\kappa$ is a scaling number. Similar dependences on both $\omega$ and $q_\|$ are seen in Figs.\,\ref{polar-2}($a$) and \ref{polar-2}($b$), respectively, where a very strong 
intrasubband-scattering resonance associated with a sign switching in ${\rm Re}[\Gamma_{n,n}(q_\|,\,\omega)]-1$ ($q_\|=q_\|^\ast$ for $n=n'=1,\,2$) occurs only within the small-value $q_\|$-$\omega$ region 
due to the presence of the $\chi^{(0)}_{n,n}(q_\|,\omega)$ interaction term in Eq.\,(\ref{eqn-7}). In this case, the intrasubband-scattering resonance is determined by the peak of

\begin{equation}
\Gamma_{n,n}(q_\|,\,\omega)=\left[1-\left(\frac{Z^\ast e^2}{2\epsilon_0\epsilon_d}\right)^2\frac{2\mu^*}{\pi^2\hbar^2}\,\chi^{(0)}_{n,n}(q_\|,\,\omega)\,
\left(\int\limits_{-{\cal L}_0/2}^{{\cal L}_0/2} dz_0\,\rho_d(z_0)\,\left|\overline{U}_{n,n}(q_\|,z_0)\right|^2\right)\right]^{-1}\ .
\end{equation}
The strength of this intrasubband-scattering resonance decreases rapidly with increasing $q_\|$ due to reduced $|\overline{U}_{n,n}(q_\|,z_0)|^2$ from the suppressed long-range intrasubband scattering 
as displayed in Fig.\,\ref{polar-1}($d$).
For intersubband excitation with $n=1$ and $n'=2$, on the other hand, the two $\Gamma_{1,2}$ terms with $q_\|$ and $q^*_\|=\sqrt{q^2_\|+8\mu^*\varepsilon_{21}/\hbar^2}$ are coupled to each other as can be 
verified by Eq.\,(\ref{eqn-7}). As a result, the broad
intersubband-scattering resonance shows up in Figs.\,\ref{polar-2}($c$) and \ref{polar-2}($d$) along with a sign switching in ${\rm Re}[\Gamma_{1,2}(q_\|,\,\omega)]-1$ and a peak in ${\rm Im}[\Gamma_{1,2}(q_\|,\,\omega)]$.
Furthermore, it is very important to notice that the broad intersubband-scattering resonance in Figs.\,\ref{polar-2}($c$) and \ref{polar-2}($d$),
due to elastic coupling between $q_\|$ and $q^\ast_\|$ electron states in two subbands,
will be different from the sharp intersubabnd-plasmon resonance determined by $\chi^{(0)}_{1,2}(q_\|,\,\omega)$ in Eq.\,(\ref{eqn-2}).
\medskip

\begin{figure}
\includegraphics[width=0.75\textwidth]{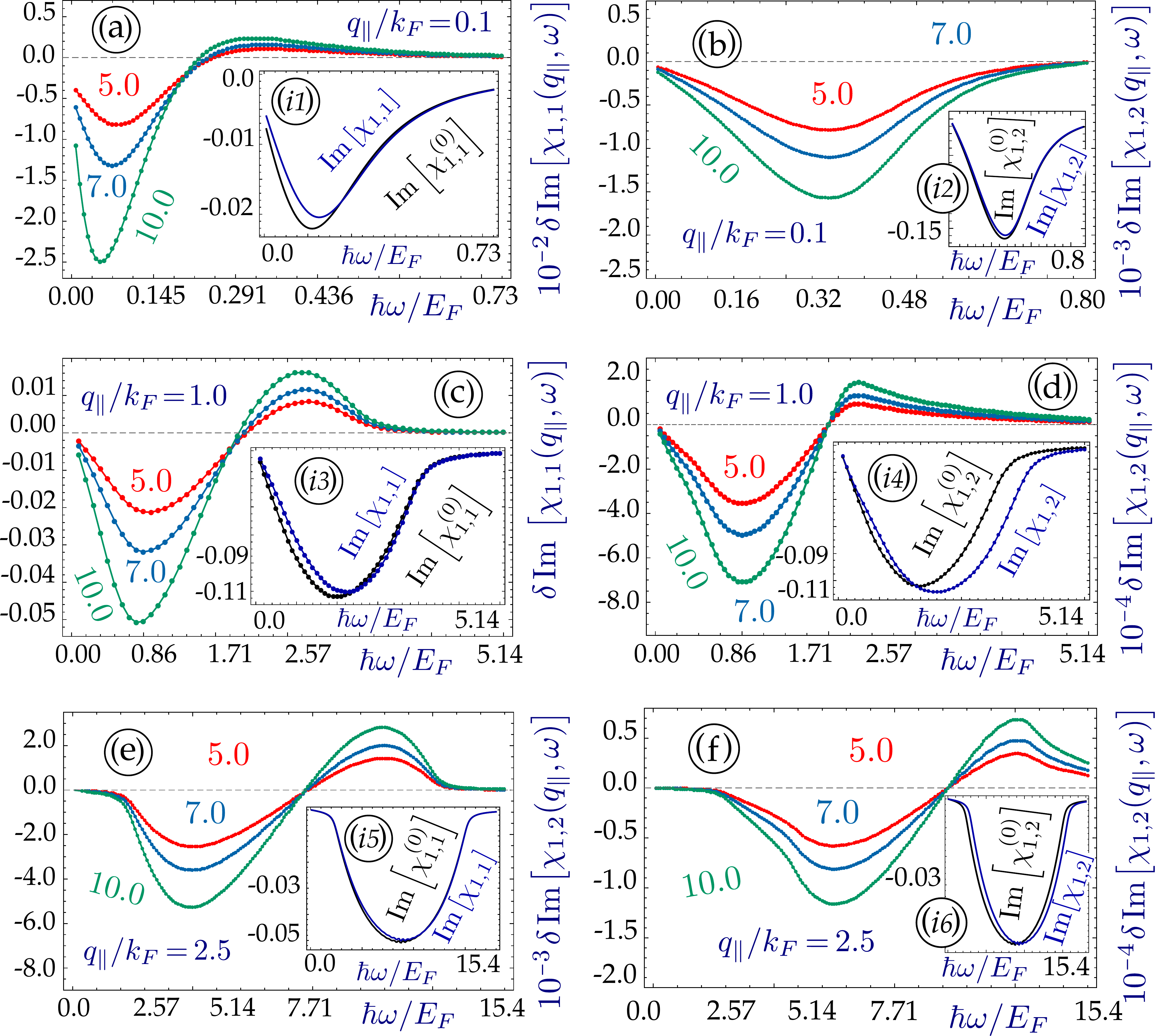}
\caption{(Color online) Defect induced variations, calculated from Eq.\,(\ref{eqn-1}), $\delta{\rm Im}[\chi_{1,1}(q_\|,\,\omega)]$ [($c$) and ($e$)] and $\delta{\rm Im}[\chi_{1,2}(q_\|,\,\omega)]$ [($d$) and ($f$)] 
for defect-density scaling number $\kappa=5,\,7,\,10$. Here, the used parameters are the same as those in Figs.\,\ref{polar-1} and \ref{polar-2}.
Results for $q_\|/k_F=0.1$, $1.0$ and $2.5$ are shown in ($a$)-($b$), ($c$)-($d$) and ($e$)-($f$), respectively. The inset of each panel compares the bare ${\rm Im}[\chi^{(0)}_{n,n'}(q_\|,\,\omega)]$ and 
screened ${\rm Im}[\chi_{n,n'}(q_\|,\,\omega)]$ in the absence of defects.
Here, the parameters used in numerical calculations are the same as those in Figs.\,\ref{polar-1} and \ref{polar-2}.}
\label{polar-3}
\end{figure}

The calculated $\Gamma_{n,n'}(q_\|,\,\omega)$ in Fig.\,\ref{polar-2} has been substituted into Eq.\,(\ref{eqn-4}) to find the intralayer dielectric function modified by defects in the RPA. 
By using Eq.\,(\ref{eqn-3}) with this modified dielectric function, the resulting inverse dielectric function 
has further been input into Eq.\,(\ref{eqn-1}) to compute related changes in the screened partial polarization functions $\delta\chi_{n,n'}(q_\|,\,\omega)$ of a single quantum well.
For intrasubband excitations in Figs.\,\ref{polar-3}($a$), \ref{polar-3}($c$) and \ref{polar-3}($e$), the defect-induced change $\delta{\rm Im}[\chi_{1,1}(q_\|,\,\omega)]$ displays a peak shift (sign switching) to a lower and lower 
value of $\omega$ with increasing $\kappa$. However, $\delta{\rm Im}[\chi_{1,1}(q_\|,\,\omega)]$ reduces significantly for a larger $q_\|$ value due to weakened scattering interaction as shown in Fig.\,\ref{polar-1}($d$). It is also interesting to notice that the depolarization shift of a plasmon peak (${\rm Im}[\chi_{1,1}(q_\|,\,\omega)]$ vs. ${\rm Im}[\chi^{(0)}_{1,1}(q_\|,\,\omega)]$) 
in the three insets of ($i1$), ($i3$) and ($i5$) (with 
$\Gamma_{n,n'}(q_\|,\,\omega)\equiv 1$) also increases with $q_\|$, but it will not show up in $\delta{\rm Im}[\chi_{1,1}(q_\|,\,\omega)]$ for defect effects. 
This pure plasmon depolarization shift to a higher $\omega$ value is rooted in a many-body screening effect and is slightly reduced by defect scatterings.
Similar features in $\delta{\rm Im}[\chi_{1,2}(q_\|,\,\omega)]$ can also be found from Figs.\,\ref{polar-3}($d$) and \ref{polar-3}($f$) 
for intersubband losses, but their magnitudes become much smaller due to very weak intersubband scattering processes.  In addition to the shift of this broad intrasubband-plasmon peak by defects, 
we also expect defect effects on a sharper intersubband-plasmon-loss peak (around $\hbar\omega\sim\varepsilon_{21}$) for a smaller $q_\|$ value, as presented in the inset of Fig.\,\ref{polar-3}($b$), 
where nearly no shift of the intensive intersubband-plasmon peak is found.

\subsection{Effects on Energy Loss of Electron Beams}
\label{sec-2-2}

In Sec\,\ref{sec-2-1}, we discussed effects of defects on the intralayer partial polarization function  $\chi_{n,n'}(q_\|,\,\omega)$.
Here, we extend our study to the kinetic-energy loss of a ballistic electron beam by further taking into account the defect effects on the interlayer total polarization function.
A full review on the excitation of collective modes, such as plasmons, in bulk materials, planar
surfaces, and nanoparticles was reported,\,\cite{gumbs-1} and the light emission induced by the electrons was proven to be an
excellent probe of plasmons, combining subnanometer resolution in the position of the electron beam
with nanometer resolution in the emitted wavelength.
\medskip

Let us assume that a semi-infinite semiconductor occupies the $z>0$ half-space and consider a classical (heavy and slow)
point charge $Q_0$ moving along a prescribed path ${\bf R}(t)$ in the air space ($z<0$) outside the semiconductor region.
In such a case, we find that the external potential $\Phi_{ext}$ associated with this moving charged particle in the quasi-static limit
satisfies the instantaneous Poisson's equation,\,\cite{EELS,gumbs} i.e.

\begin{equation}
\nabla^2_{\bf r}\Phi_{ext}({\bf r},t\vert{\bf R})=-\frac{Q_0}{\epsilon_0}\,\delta\left[{\bf r}-{\bf R}(t)\right]\ ,
\label{eqn-9}
\end{equation}
where ${\bf R}(t)=\{{\bf R}_\|(t),\,Z(t)\}$ is the trajectory of the charged particle, and ${\bf r}=\{{\bf r}_\|,\,z\}$ is a position vector.
The solution of Eq.\,(\ref{eqn-9}) inside the region of $Z(t)<z<0$ is found to be

\begin{equation}
\Phi^<_{ext}({\bf r},t\vert{\bf R})=\int \frac{d^2{\bf q}_\|}{(2\pi)^2}\int\limits_{-\infty}^{\infty}\frac{d\omega}{2\pi}\,\phi_{ext}({\bf q}_\|,\omega\vert{\bf R})\,e^{i{\bf q}_\|\cdot{\bf r}_\|-i\omega t}\,e^{-q_\|z}\ ,
\label{eqn-10}
\end{equation}
where the Fourier-transformed external potential is calculated as

\begin{equation}
\phi_{ext}({\bf q}_\|,\omega\vert{\bf R})=-\frac{Q_0}{2\epsilon_0q_\|}\,{\cal F}_0({\bf q}_\|,\omega\vert{\bf R})\ ,
\label{eqn-11}
\end{equation}
and its structure factor is

\begin{equation}
{\cal F}_0({\bf q}_\|,\omega\vert{\bf R})=\int\limits_{-\infty}^{\infty} dt'\,e^{q_\|Z(t')}\,e^{i\omega t'-i{\bf q}_\|\cdot{\bf R}_\|(t')}\ .
\label{eqn-12}
\end{equation}
\medskip

From a physics perspective, the existence of $\Phi_{ext}$ inside the semiconductor will induce a potential $\Phi_{ind}$ outside the semiconductor (i.e., $z<0$)
due to the charge-density fluctuation, yielding

\begin{equation}
\Phi^<_{ind}({\bf r},t\vert{\bf R})=-\int \frac{d^2{\bf q}_\|}{(2\pi)^2}\int\limits_{-\infty}^{\infty}\frac{d\omega}{2\pi}\,\phi_{ext}({\bf q}_\|,\omega\vert{\bf R})\,e^{i{\bf q}_\|\cdot{\bf r}_\|-i\omega t}\,
{\cal S}(q_\|,\,\omega)\,e^{q_\|z}\ ,
\label{eqn-13}
\end{equation}
where ${\cal S}(q_\|,\,\omega)$ is the so-called surface-response function\,\cite{book} determined later by matching the boundary condition.
Within the semiconductor region ($0\leq z\leq {\cal L}_0$), we write down similar expressions for the external $\Phi^>_{ext}$ and induced $\Phi^>_{ind}$ potentials, given by

\begin{equation}
\Phi^>_{ext}({\bf r},t\vert{\bf R})=\int \frac{d^2{\bf q}_\|}{(2\pi)^2}\int\limits_{-\infty}^{\infty}\frac{d\omega}{2\pi}\,\phi_{ext}({\bf q}_\|,\omega\vert{\bf R})\,
e^{i{\bf q}_\|\cdot{\bf r}_\|-i\omega t}\,\Phi^>_0(z\vert q_\|)\ ,
\label{eqn-14}
\end{equation}

\begin{equation}
\Phi^>_{ind}({\bf r},t\vert{\bf R})=-\int \frac{d^2{\bf q}_\|}{(2\pi)^2}\int\limits_{-\infty}^{\infty}\frac{d\omega}{2\pi}\,\phi_{ext}({\bf q}_\|,\omega\vert{\bf R})\,e^{i{\bf q}_\|\cdot{\bf r}_\|-i\omega t}\,
\phi^{>}_{ind}(z\vert q_\|,\omega)\ ,
\label{eqn-15}
\end{equation}
where $\Phi^>_0(z\vert q_\|)$ is the bare external potential in the electrostatic limit ($q_\| c\gg\omega$) for a slab of semiconductor material of thickness ${\cal L}_0$, $\Phi^>_0(0\vert q_\|)=1-g_{slab}(q_\|)$,
and $g_{slab}(q_\|)$ is the surface-response function for a dielectric slab without doping electrons.\,\cite{book}
Since the total potential $\Phi^>_0(z\vert q_\|)+\phi^{>}_{ind}(z\vert q_\|,\omega)$ inside the semiconductor ($z>0$) equals the screened external potential,
we get $\phi^{>}_{ind}$ in Eq.\,(\ref{eqn-15}) from\,\cite{nrc}

\begin{equation}
\phi^{>}_{ind}(z\vert q_\|,\omega)=\int dz'\left[\epsilon^{-1}(z,z'\vert q_\|,\omega)-\delta(z-z')\right]\Phi^>_0(z'\vert q_\|)\ .
\label{eqn-16}
\end{equation}
In Eq.\,(\ref{eqn-16}), the inverse dielectric function can be found from

\begin{equation}
\epsilon^{-1}(z,z'\vert q_\|,\omega)=\delta(z-z')+\int dz^{\prime\prime}\,V_{c}(z,z^{\prime\prime}\vert q_\|)\,\chi(z^{\prime\prime},z'\vert q_\|,\omega)\ ,
\label{eqn-17}
\end{equation}
where the interlayer Coulomb coupling $V_{c}(z,z'\vert q_\|)$, including the image potentials, is calculated as\,\cite{book}

\begin{equation}
V_{c}(z,z'\vert q_\|)=\frac{\beta_0(q_\|)\,e^2}{2\epsilon_0\epsilon_{d}(q_\|+q_s)}\left[e^{-q_\||z-z'|}+\alpha_0^2\,e^{-2q_\|{\cal L}_0}\,e^{q_\||z-z'|}+\alpha_0\,e^{-q_\||z+z'|}
+\alpha_0\,e^{-2q_\|{\cal L}_0}\,e^{q_\||z+z'|}\right]\ ,
\label{eqn-18}
\end{equation}
and $\alpha_0=(\epsilon_{d}-1)/(\epsilon_{d}+1)$, $\beta_0(q_\|)=1/[1-\alpha_0^2\exp(-2q_\|{\cal L}_0)]$.
\medskip

For a multi-quantum-well system, the density-density-response function in Eq.\,(\ref{eqn-17}) takes the form\,\cite{EELS}

\begin{equation}
\chi(z,z'\vert q_\|,\omega)=\sum_{j,j'=0}^N\,\delta(z-ja)\,\tilde{\chi}_e(j,j'\vert q_\|,\omega)\,\delta(z'-j'a)\ ,
\label{eqn-19}
\end{equation}
where $a$ is the well separation, ${\cal L}_0=Na$,
and the screened polarization function $\tilde{\chi}_e(j,j'\vert q_\|,\omega)$ within the RPA
can be obtained from the following self-consistent equations\,\cite{EELS}

\begin{equation}
\tilde{\chi}_e(j,j'\vert q_\|,\omega)=\tilde{\chi}_j(q_\|,\,\omega)\,\delta_{j,j'}+\tilde{\chi}_j(q_\|,\,\omega)\sum_{j^{\prime\prime}(\neq j)=0}^N\,V_{c}(ja,j^{\prime\prime}a\vert q_\|)\,\tilde{\chi}_e(j^{\prime\prime},j'\vert q_\|,\omega)\ .
\label{eqn-20}
\end{equation}
Here, the summation over $j^{\prime\prime}$ excludes the intralayer term with $j^{\prime\prime}=j$, the integers $j=0,\,1,\,\cdots,\,N$ labels different wells,
and $\displaystyle{\tilde{\chi}_j(q_\|,\,\omega)=\sum_{n\leq n'}\,\chi_{n,n'}(j,j\vert q_\|,\omega)}$ is the total
polarization function for the $j$th quantum well as discussed in Sec.\,\ref{sec-2-1}.
\medskip

By combining Eqs.\,(\ref{eqn-16}), (\ref{eqn-17}) and (\ref{eqn-19}), $\phi^{>}_{ind}(z\vert q_\|,\omega)$ in Eq.\,(\ref{eqn-15}) can be simply rewritten as

\begin{equation}
\phi^{>}_{ind}(z\vert q_\|,\omega)=\sum_{j,j'=0}^N\,V_{c}(z,ja\vert q_\|)\,\tilde{\chi}_e(j,j'\vert q_\|,\omega)\,\Phi^>_0(j'a\vert q_\|)\ .
\label{eqn-21}
\end{equation}
By matching the boundary condition for the total potential, i.e., $1-{\cal S}(q_\|,\,\omega)=[1-g_{slab}(q_\|)]+\phi^{>}_{ind}(0\vert q_\|,\omega)$ at the surface $z=0$, we are able to
find the surface response function introduced in Eq.\,(\ref{eqn-13}) from

\begin{equation}
{\cal S}(q_\|,\,\omega)=g_{slab}(q_\|)-\sum_{j,j'=0}^N\,V_{c}(0,ja\vert q_\|)\,\tilde{\chi}_e(j,j'\vert q_\|,\omega)\,\Phi^>_0(j'a\vert q_\|)\ ,
\label{eqn-22}
\end{equation}
where\,\cite{EELS}

\begin{equation}
g_{slab}(q_\|)=2\alpha_0\,\beta_0(q_\|)\,e^{-q_\| Na}\,\sinh(q_\| Na)\ ,
\label{eqn-23}
\end{equation}
and the external electrostatic potential in Eqs.\,(\ref{eqn-16}) and (\ref{eqn-22}) inside a slab of semiconductor ($0\leq z\leq Na$) is found to be\,\cite{EELS}

\begin{equation}
\Phi^>_0(z\vert q_\|)=\left[\frac{1-g_{slab}(q_\|)}{2}+\frac{1+g_{slab}(q_\|)}{2\epsilon_{d}}\right]e^{-q_\| z}
+\left[\frac{1-g_{slab}(q_\|)}{2}-\frac{1+g_{slab}(q_\|)}{2\epsilon_{d}}\right]e^{q_\| z}\ .
\label{eqn-24}
\end{equation}
\medskip

The absorbed kinetic energy $\Delta E_{abs}\{{\bf R}\}$ of an electron beam
can be calculated by integrating the Poynting vector over the surface and over time in the air region, which leads to\,\cite{nrc}

\begin{equation}
\Delta E_{abs}\{{\bf R}\}=\epsilon_0\int d^2{\bf r}_\|\int\limits_{-\infty}^{\infty} dt\,{\rm Re}\left.\left\{\left[\Phi^<_{tot}({\bf r},t\vert{\bf R})\right]^\ast
\frac{\partial^2\Phi^<_{tot}({\bf r},t\vert{\bf R})}{\partial t\,\partial z}\right\}\right|_{z=0}\ ,
\label{eqn-25}
\end{equation}
where $\Phi^<_{tot}({\bf r},t\vert{\bf R})$ is the total potential outside the semiconductor region ($z<0$), calculated by combining Eqs.\,(\ref{eqn-10}) and (\ref{eqn-13}) and given by

\begin{equation}
\Phi^<_{tot}({\bf r},t\vert{\bf R})=\int \frac{d^2{\bf q}_\|}{(2\pi)^2}\int\limits_{-\infty}^{\infty}\frac{d\omega}{2\pi}\,\left[e^{-q_\|z}-{\cal S}(q_\|,\,\omega)\,e^{q_\|z}\right]\,
e^{i{\bf q}_\|\cdot{\bf r}_\|-i\omega t}\,\phi_{ext}({\bf q}_\|,\omega\vert{\bf R})\ .
\label{eqn-26}
\end{equation}
Substituting this result into Eq.\,(\ref{eqn-25}), we find

\begin{equation}
\Delta E_{abs}\{{\bf R}\}=\frac{Q_0^2}{2\epsilon_0}\int \frac{d^2{\bf q}_\|}{(2\pi)^2}\int\limits_{-\infty}^{\infty}\frac{d\omega}{2\pi}\left(\frac{\left|{\cal F}_0({\bf q}_\|,\omega\vert{\bf R})\right|^2\omega}{q_\|}\right)
{\rm Im}\left\{{\cal S}(q_\|,\,\omega)\right\}\ ,
\label{eqn-27}
\end{equation}
where ${\rm Im}\left\{{\cal S}(q_\|,\,\omega)\right\}$ is the so-called loss function.\,\cite{book}
\medskip

Specifically, for a charged particle moving parallel to the surface, we have ${\bf R}(t)=\{{\bf V}_\|t,\,Z_0\}$ and obtain

\begin{equation}
\left|{\cal F}_0({\bf q}_\|,\omega\vert{\bf R})\right|^2=\lim_{\Delta T\to\infty}\left|\int\limits_{-\Delta T/2}^{\Delta T/2} dt'\,e^{-q_\|Z_0}\,e^{i(\omega-{\bf q}_\|\cdot{\bf V}_\|)t'}\right|^2
=2\pi\Delta T\,e^{-2q_\| |Z_0|}\,\delta(\omega-{\bf q}_\|\cdot{\bf V}_\|)\ ,
\label{eqn-28}
\end{equation}
which leads to the following power absorption for the parallel electron beam

\begin{equation}
\frac{\Delta E_{abs}(V_\|)}{\Delta T}=\frac{Q_0^2}{2\epsilon_0}\int \frac{d^2{\bf q}_\|}{(2\pi)^2}\,e^{-2q_\| |Z_0|}\,\left(\frac{{\bf q}_\|\cdot{\bf V}_\|}{q_\|}\right)\,
{\rm Im}\left[{\cal S}(q_\|,\,{\bf q}_\|\cdot{\bf V}_\|)\right]\ .
\label{eqn-29}
\end{equation}
More interesting, if a charged particle moves away from the surface perpendicularly, we can write ${\bf R}(t)=\{0,\,Z_0-V_\perp t\}$ with an impact parameter $|Z_0|$ ($Z_0<0$) and $0\leq t\leq T_0$ for the damped particle,
and obtain

\[
\left|{\cal F}_0({\bf q}_\|,\omega\vert{\bf R})\right|^2=\lim_{T_0\to\infty}\left|\int\limits_{0}^{T_0} dt'\,e^{-q_\|(|Z_0|+V_\perp t')}\,e^{i\omega t'}\right|^2
\]
\begin{equation}
=\lim_{T_0\to\infty}\left|\frac{e^{-q_\| |Z_0|}}{q_\|V_\perp-i\omega}\,\left[1-e^{(q_\|V_\perp-i\omega)T_0}\right]\right|^2
=\frac{e^{-2q_\| |Z_0|}}{\omega^2+q_\|^2V_\perp^2}\ ,
\label{eqn-30}
\end{equation}
which yields the energy absorption for the perpendicular electron beam

\begin{equation}
\Delta E_{abs}(V_\perp)=\frac{Q_0^2}{2\epsilon_0}\int \frac{d^2{\bf q}_\|}{(2\pi)^2}\int \frac{d\omega}{2\pi}\,\left(\frac{\omega}{q_\|}\right)\,\frac{e^{-2q_\| |Z_0|}}{\omega^2+q_\|^2V_\perp^2}\,
{\rm Im}\left\{{\cal S}(q_\|,\,\omega)\right\}\ .
\label{eqn-31}
\end{equation}
In this case, the integral over $\omega$ with respect to the loss function ${\rm Im}\{{\cal S}(q_\|,\,\omega)\}$ includes the damping contributions from both the particle-hole and collective excitation
modes of electrons.\,\cite{add-15}
\medskip

Multiple plasmon excitations in graphene materials by a single electron was predicted to give rise to a unique
platform for exploring the bosonic quantum nature of these collective modes.\,\cite{gumbs-2} Such a technique not only
opens a viable path toward multiple excitation of a single plasmon mode by a single electron, but also reveals electron probes as ideal tools
for producing, detecting, and manipulating plasmons in graphene nanostructures.
\medskip

\begin{figure}
\includegraphics[width=0.75\textwidth]{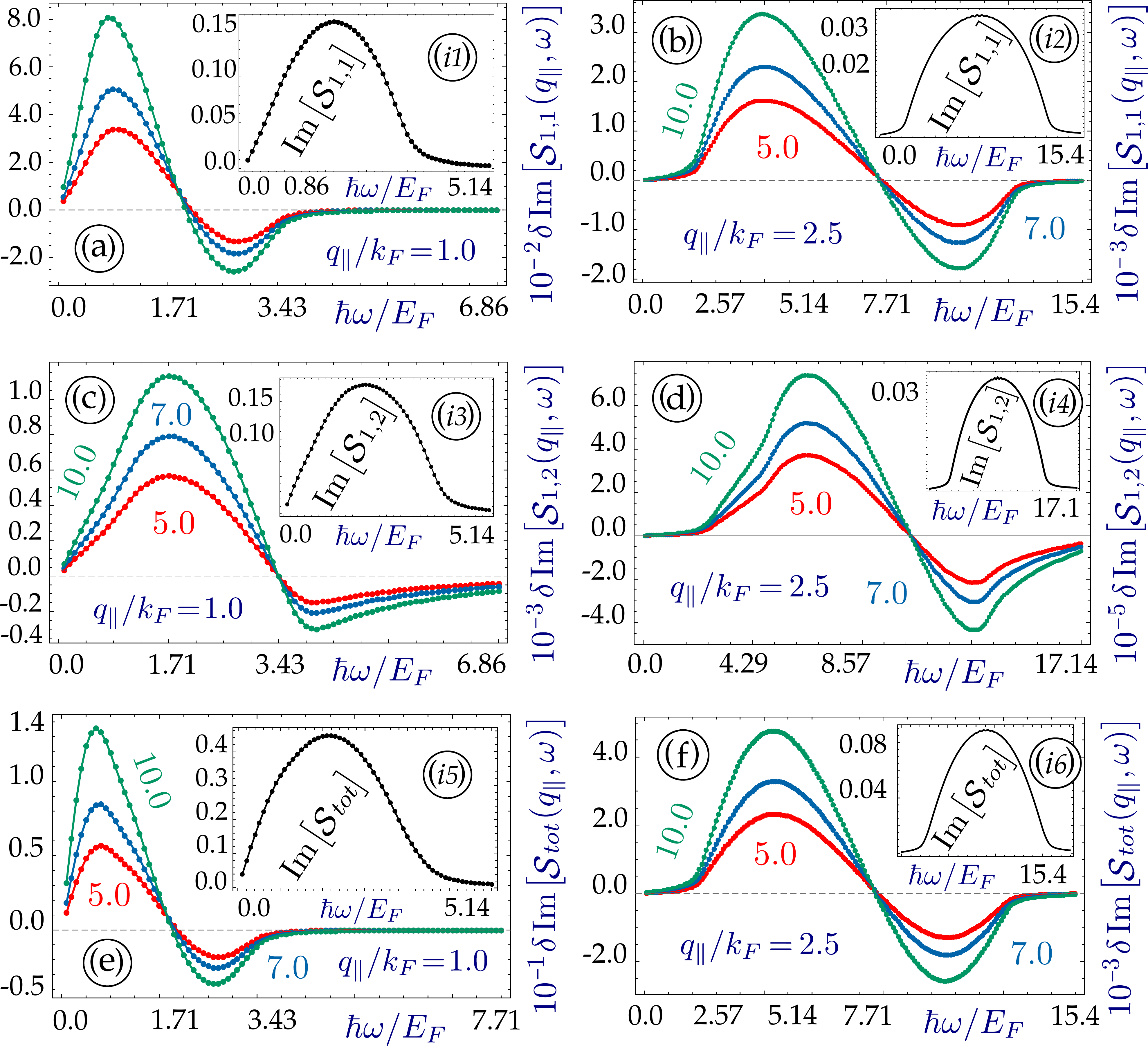}
\caption{(Color online) Defect induced variations in single-quantum-well partial and total loss functions $\displaystyle{\delta{\rm Im}[{\cal S}(q_\|,\,\omega)]=\sum_{n\leq n'}\,\delta{\rm Im}[{\cal S}_{n,n'}(q_\|,\,\omega)]}$ 
calculated from Eq.\,(\ref{eqn-22}) for $q_\|/k_F=1.0$ [($a$), ($c$), ($e$)] 
and $q_\|/k_F=2.5$ [($b$), ($d$), ($f$)] with different defect-density scaling numbers $\kappa=5,\,7,\,10$. 
The inset of each panel displays partial ${\rm Im}[{\cal S}_{n,n'}(q_\|,\,\omega)]$ and 
total ${\rm Im}[{\cal S}(q_\|,\,\omega)]$ in the absence of defects.
Here, the parameters used in numerical calculations are the same as those in Figs.\,\ref{polar-1} and \ref{polar-2}.}
\label{polar-4}
\end{figure}

For a single quantum well, the surface response function $\displaystyle{{\cal S}(q_\|,\,\omega)=\sum_{n\leq n'}\,{\cal S}_{n,n'}(q_\|,\omega)}$ can be obtained by setting $j=j'=0$ in Eq.\,(\ref{eqn-22}),
and the total loss function is just $\displaystyle{{\rm Im}[{\cal S}(q_\|,\,\omega)]=\sum_{n\leq n'}\,{\rm Im}[{\cal S}_{n,n'}(q_\|,\omega)]}$. Here, the defect induced change  
$\delta{\rm Im}[{\cal S}_{n,n'}(q_\|,\omega)]$ directly relates to 
the imaginary part of the screened partial polarization function $\delta{\rm Im}[\chi_{n,n'}(q_\|,\omega)]$ presented in Fig.\,\ref{polar-3}. For $q_\|/k_F=1.0$, we find from  
Figs.\,\ref{polar-4}($a$), \ref{polar-4}($c$) and \ref{polar-4}($e$) that $\delta{\rm Im}[{\cal S}(q_\|,\,\omega)]$ is dominated by $\delta{\rm Im}[{\cal S}_{1,1}(q_\|,\,\omega)]$ for a stronger 
intrasubband scattering process, which increases with the defect-density scaling number $\kappa$. The sign switching reflects the shift of a loss peak [see insets of 
Figs.\,\ref{polar-4}($a$), \ref{polar-4}($c$) and \ref{polar-4}($e$)] to a lower value of $\omega$. 
As $q_\|/k_F$ is increased to $2.5$ in Figs.\,\ref{polar-4}($b$), \ref{polar-4}($d$) and \ref{polar-4}($f$), the resonant peak of ${\rm Im}[{\cal S}(q_\|,\,\omega)]$ moves to a higher 
$\omega$ value [see insets ($i5$) and ($i6$)]. However, the similar defect-related features as in Figs.\,\ref{polar-4}($a$), \ref{polar-4}($c$) and \ref{polar-4}($e$)
are greatly weakened due to a dramatic reduction of scattering interactions as shown in Fig.\,\ref{polar-1}($d$).
\medskip

\begin{figure}
\includegraphics[width=0.75\textwidth]{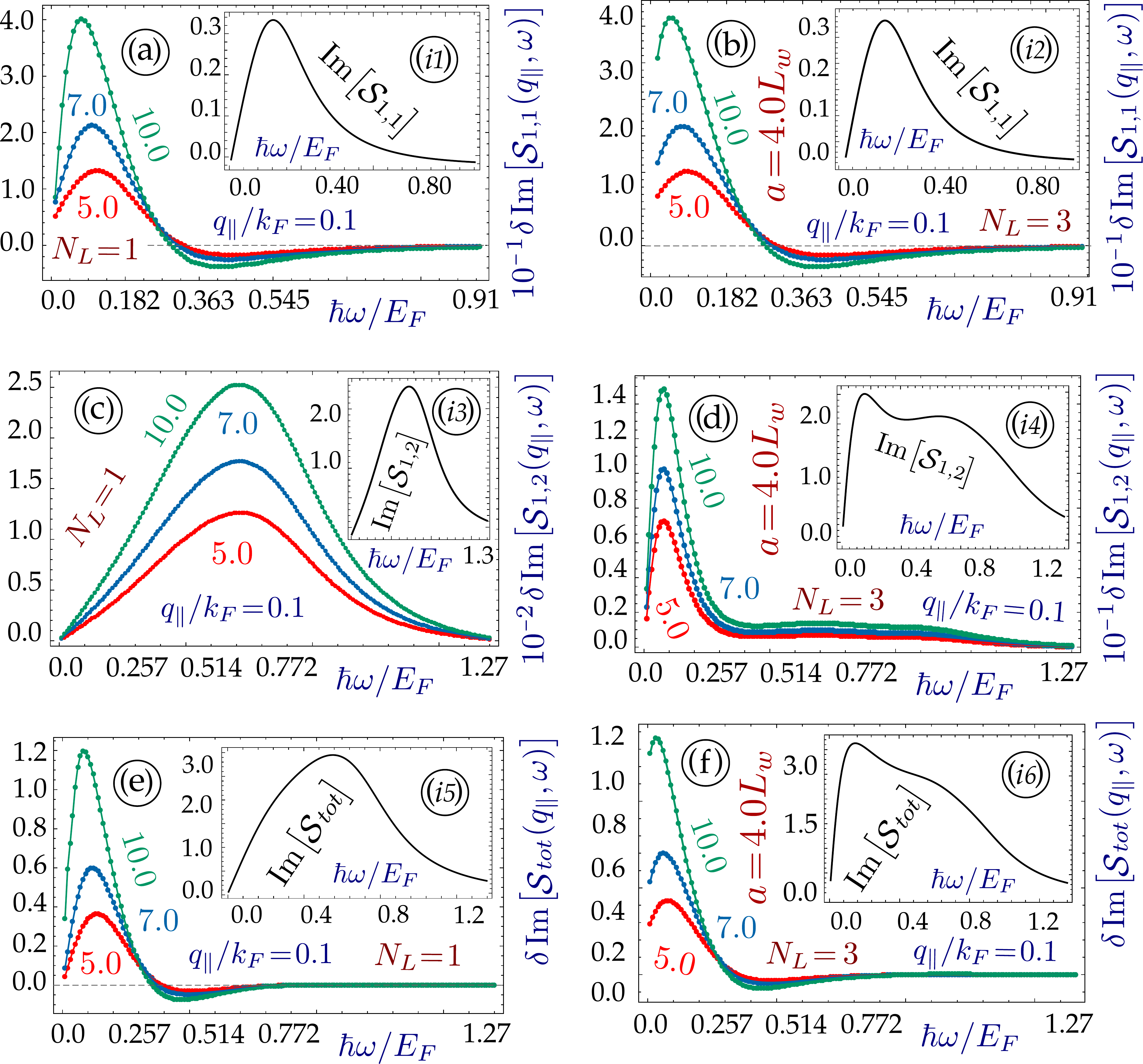}
\caption{(Color online) Comparisons of defect induced variations in single- and multi-quantum-well partial and total loss functions 
$\displaystyle{\delta{\rm Im}[{\cal S}(q_\|,\,\omega)]=\sum_{n\leq n'}\,\delta{\rm Im}[{\cal S}_{n,n'}(q_\|,\,\omega)]}$ calculated from Eq.\,(\ref{eqn-22}) at $q_\|/k_F=0.1$ 
for the numbers of quantum wells $N_L=1$ [($a$), ($c$), ($e$)] 
and $N_L=3$ [($b$), ($d$), ($f$)] with different defect-density scaling numbers $\kappa=5,\,7,\,10$. 
For defect distribution, we still use $\rho_d(z_0)/\kappa=\rho_1\Theta(-z_0-L_W/2)]+\rho_2\Theta(z_0-L_W/2)+[\rho_0+z_0(\Delta\rho/L_W)]\Theta(L_W/2-|z_0|)$ for each quantum well and two outer barriers, 
while $\rho_d(z_0)/\kappa$ is set to $\rho_2$ for the regions between two adjacent quantum wells.   
The inset of each panel displays partial ${\rm Im}[{\cal S}_{n,n'}(q_\|,\,\omega)]$ and 
total ${\rm Im}[{\cal S}(q_\|,\,\omega)]$ in the absence of defects.
Here, $a/L_W=4$ in ($b$), ($d$), ($f$)
and the other parameters used in numerical calculations are the same as those in Figs.\,\ref{polar-1} and \ref{polar-2}.}
\label{polar-5}
\end{figure}

For a multi-quantum well system, the interlayer Coulomb coupling $V_c(ja,j'a\vert q_\|)$ in Eq.\,(\ref{eqn-20}) will modify the intralayer total polarization function $\tilde{\chi}_j(q_\|,\,\omega)$, as well
as the surface response function in Eq.\,(\ref{eqn-22}). From the comparison of single- and multi-quantum well systems in Fig.\,\ref{polar-5}, we find the intersubband-plasmon loss ${\rm Im}[{\cal S}_{1,2}(q_\|,\,\omega)]$ 
is strongly coupled to the intrasubband-plasmon loss ${\rm Im}[{\cal S}_{1,1}(q_\|,\,\omega)]$ by interlayer Coulomb coupling, as shown in the insets of ($i4$) and ($i6$).
Here, the weaker ${\rm Im}[{\cal S}_{1,1}(q_\|,\,\omega)]$ peak in the inset ($i2$) is greatly enhanced by its sitting on the shoulder of a much stronger ${\rm Im}[{\cal S}_{1,2}(q_\|,\,\omega)]$ peak in the inset ($i4$), 
giving rise to a profile for the total ${\rm Im}[{\cal S}(q_\|,\,\omega)]$ peak in the inset ($i6$).
As $q_\|/k_F=0.1$, the defect-induced peak shift in $\delta{\rm Im}[{\cal S}_{1,1}(q_\|,\,\omega)]$ to lower $\omega$ can be seen from Fig.\,\ref{polar-5}($b$) but not for $\delta{\rm Im}[{\cal S}_{1,2}(q_\|,\,\omega)]$ 
in Fig.\,\ref{polar-5}($d$) except for a significant enhancement of the shoulder peak with increasing $\kappa$ by interlayer Coulomb coupling. 
Moreover, by comparing Figs.\,\ref{polar-5}($a$) with \ref{polar-5}($b$), we find both ${\rm Im}[{\cal S}_{1,1}(q_\|,\,\omega)]$ and $\delta{\rm Im}[{\cal S}_{1,1}(q_\|,\,\omega)]$ are dominated by the 
intralayer Coulomb coupling ${\cal V}_{m,m';n,n'}(q_\|)$ given by Eq.\,(\ref{eqn-5}).

\subsection{Effects on Loss of Photons}
\label{sec-2-3}

In Sec.\,\ref{sec-2-2}, the defect effects on the energy loss of electron beams in a multi-quantum-well system was discussed. As a comparison, the defect effects on the loss of photons
(or photon absorption) in the same system will be investigated here. In this case, both the absorption coefficients for intrasubband and intersubband optical transitions of electrons can be calculated from\,\cite{talwar}

\begin{equation}
\beta_{abs}(\omega)=\frac{\epsilon_{d}\,\omega}{n_{r}(\omega)c}\,\left[1+\frac{1}{\exp(\hbar\omega/k_BT)-1}\right]\,{\rm Im}\{\alpha_{L}(\omega)\}\ ,
\label{eqn-32}
\end{equation}
where $\hbar\omega$ is the incident-photon energy, and the dynamical refractive-index function $n_{r}(\omega)$ is

\begin{equation}
n_{r}(\omega)=\sqrt{\frac{\epsilon_{d}}{2}}\left\{1+{\rm Re}\{\alpha_{L}(\omega)\}+\sqrt{[1+{\rm Re}\{\alpha_{L}(\omega)]^2+[{\rm Im}\{\alpha_{L}(\omega)]^2}\right\}^{1/2}\ .
\label{eqn-33}
\end{equation}
\medskip

For intrasubband transitions with an optical probe field polarized parallel to the quantum-well planes,
$\alpha_{L}(\omega)$ in Eqs.\,(\ref{eqn-32}) and (\ref{eqn-33}) is the Lorentz ratio calculated as\,\cite{add-1}

\begin{equation}
\alpha_{L}(\omega)=\alpha^\|_{L}(\omega)=-\left(\frac{2e^2}{\epsilon_0\epsilon_{d}L_{W}}\right)\pi{\cal R}_0^2\,\sum_{j=0}^N\,
\int \frac{d^2{\bf q}_\|}{(2\pi)^2}\,e^{-q^2_\|{\cal R}_0^2/4}\,{\cal Q}^\|_j(q_\|,\,\omega)\ ,
\label{eqn-34}
\end{equation}
where ${\cal R}_0$ is the radius of a normally-incident Gaussian light beam, $N+1$ is the total number of quantum wells in the system, and
the optical-response function\,\cite{add-16} ${\cal Q}^\|_j(q_\|,\,\omega)$ for the $j$th well is found to be

\begin{equation}
{\cal Q}^\|_j(q_\|,\,\omega)=\sum_{n}\,\tilde{\chi}_{n,n}(j,j\vert q_\|,\omega)\,\left(\frac{\hbar q_\|}{\mu^\ast\omega}\right)^2\ .
\label{eqn-35}
\end{equation}
\medskip

By including the couping due to interlayer Coulomb interactions, the partial polarization function $\tilde{\chi}_{n,n}(j,j';\,q_\|,\omega)$
introduced in Eq.\,(\ref{eqn-35}) with $j=j'$ needs to be computed from the following self-consistent equations\,\cite{EELS} (taking $n=n'$ and $j=j'$ afterwards), i.e.,

\begin{equation}
\tilde{\chi}_{n,n'}(j,j'\vert q_\|,\omega)=\chi_{n,n'}(q_\|,\,\omega)\,\delta_{j,j'}+\chi_{n,n'}(q_\|,\,\omega)
\sum_{j^{\prime\prime}(\neq j)=0}^N\,
V_{c}(ja,j^{\prime\prime}a\vert q_\|)\,\tilde{\chi}_{n,n'}(j^{\prime\prime},j'\vert q_\|,\omega)\ ,
\label{eqn-36}
\end{equation}
where $\chi_{n,n'}(q_\|,\,\omega)\equiv\tilde{\chi}_{n,n'}(j,j\vert q_\|,\omega)$, and the interlayer Coulomb matrix elements $V_{c}(ja,j^{\prime\prime}a\vert q_\|)$ are still found from Eq.\,(\ref{eqn-18}).
By further taking into account the coupling between different subbands in each quantum well,
the screened partial polarization function $\chi_{n,n'}(q_\|,\,\omega)$ in Eq.\,(\ref{eqn-36})
must be calculated from Eq.\,(\ref{eqn-1}) after finding the inverse dielectric function from Eqs.\,(\ref{eqn-3}) and (\ref{eqn-4}).
\medskip

On the other hand, for a spatially-uniform optical probe field polarized perpendicular to the quantum-well planes, the Lorentz ratio $\alpha_{L}(\omega)$ in Eqs.\,(\ref{eqn-32}) and (\ref{eqn-33})
for intersubband transitions becomes\,\cite{add-1}

\begin{equation}
\alpha_{L}(\omega)=\alpha^\perp_{L}(\omega)=-\frac{2e^2}{\epsilon_0\epsilon_{d}L_{W}}\,
\sum_{j=0}^N\,{\cal Q}^\perp_j(q_\|=0,\,\omega)\ ,
\label{eqn-37}
\end{equation}
where we have assumed $q_\|/k_{F}=\sqrt{\epsilon_{d}}\,\omega/k_{F}c\ll 1$, and $k_{F}$ is the Fermi wavenumber of electrons in quantum wells.
In this case, the optical-response function for the $j$th well in Eq.\,(\ref{eqn-37}) takes the form\,\cite{add-16}

\begin{equation}
{\cal Q}^\perp_j(q_\|,\,\omega)=\sum_{n<n'}\,\tilde{\chi}_{n,n'}(j,j\vert q_\|,\omega)\,\left|\int_{-\infty}^{\infty} dz\,{\cal F}_{n'}(z)\,z\,{\cal F}_n(z)\right|^2\ .
\label{eqn-38}
\end{equation}
Moreover, the influence of interlayer Coulomb coupling on the intersubband partial polarization function $\tilde{\chi}_{n,n'}(j,j'\vert q_\|,\omega)$ should still be determined from Eq.\,(\ref{eqn-36})
(setting $j=j'$ afterwards).
\medskip

A periodic stack of graphene layers is expected to have the properties of
a one-dimensional photonic crystal with stop bands at certain frequencies. As an
incident electromagnetic wave is reflected from these stacked graphene layers, the tuning of the graphene Fermi energy or 
conductivity renders the possibility of controlling these stop bands, leading to a tunable spectral-selective mirror.\,\cite{gumbs-3}
In addition, a transfer-matrix method was applied to explore optical reflection, transmission and absorption in
single-, double- and multi-layer graphene structures.\,\cite{gumbs-4} Both the total internal reflection in single-layer
graphene, as well as thin-film interference effects in double-layer graphene, are shown for increasing light absorption.
\medskip

\begin{figure}
\includegraphics[width=0.75\textwidth]{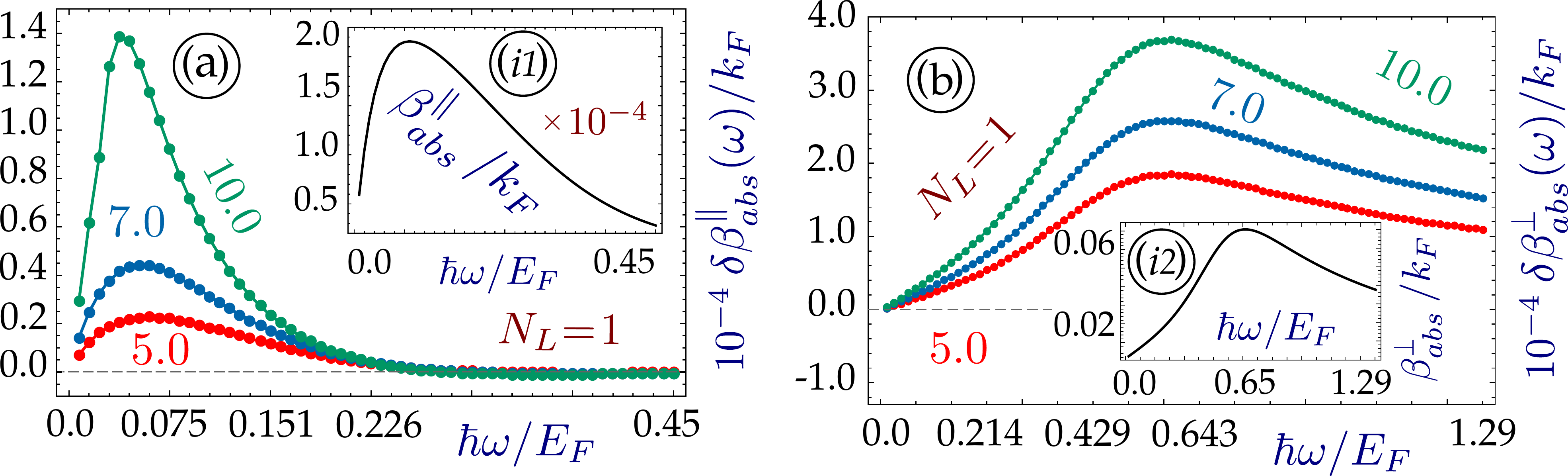}
\caption{(Color online) Defect induced variations in single-quantum-well intrasubband $\delta\beta^\|_{abs}(\omega)$ and intersubband $\delta\beta^\perp_{abs}(\omega)$ absorption coefficients (in units of $k_F$),
calculated respectively from Eqs.\,(\ref{eqn-34}) and (\ref{eqn-37}), for $N_L=1$ and different defect-density scaling numbers $\kappa=5,\,7,\,10$. 
The insets ($i1$) and ($i2$) present $\beta^\|_{abs}(\omega)$ and $\beta^\perp_{abs}(\omega)$ in the absence of defects.
Here, $k_F{\cal R}_0=50$ and the other parameters used in numerical calculations are the same as those in Figs.\,\ref{polar-1} and \ref{polar-2}.}
\label{polar-6}
\end{figure}

For intrasubband electron transitions induced by an optical field with a polarization parallel to the quantum-well plane, we present in Fig.\,\ref{polar-6}($a$) the defect modification to the absorption coefficient 
$\delta\beta^\|_{abs}(\omega)$ calculated from Eqs.\,(\ref{eqn-32}) and (\ref{eqn-34}). Here, the low-energy photon absorption peak in the inset ($i1$) is attributed to the excitation of intrasubband plasmons, and 
this peak is shifted to an even lower $\omega$ value with increasing $\kappa$. On the other hand, for the intersubband transition of electrons under an optical field polarized perpendicular to the quantum-well plane, 
we display in Fig.\,\ref{polar-6}($b$) the defect changes in absorption coefficient $\delta\beta^\perp_{abs}(\omega)$ calculated from Eqs.\,(\ref{eqn-32}) and (\ref{eqn-37}). 
In this case, however, a high-energy and broad photon absorption peak in the inset ($i2$) results from intrasubband-plasmon excitations, and no shift associated with this peak with $\kappa$ is found.

\section{Ultrafast Point-Defect Dynamics}
\label{sec-3}

In Sec.\,\ref{sec-2}, we only discuss the effects of point defects on losses of electron energy and photons in a multi-quantum-well system.
In Sec.\,\ref{sec-3}, we explore ultrafast dynamics for the production of Frenkel-pair defects and their follow-up reactions and diffusions in the same system.
In this way, the spatial dependence of the one-dimensional distribution function $\rho_d(z)$ introduced in Eq.\,(\ref{eqn-7}) for the defect-electron interaction can be extracted.
It is known that the Frenkel-pair production will be followed subsequently by diffusion and reactions to reach defect stabilization through diffusion-induced recombination and reactions with
residual defects in the system. Here, the diffusion of point defects is driven by forces other than the concentration gradient of defects, 
e.g., compressive stress near sinks. The reactions, on the other hand, are enabled by the presence of
growth-induced dislocation loops at the two interfaces of a quantum well.

\subsection{Defect Diffusion-Reaction Equations}
\label{sec-3-1}

Let us start by considering an $N$ layered material structure in the $z$ direction.
Each material layer is characterized by the (bulk) irradiation parameters ${\cal G}_0^j$, ${\cal R}^j$, $D^j$ and $\Gamma^j(t)$
with layer labels $j=1,\,2,\,3,\,\cdots,\,N$
for production and recombination rates, diffusion coefficient and bulk-sink annihilation, respectively.
In modeling a mesoscopic-scale sample, the interface-sink strengths $[\kappa^j(t)]^2$ with $j=1,\,2,\,3,\,\cdots,\,N-1$ also need to be taken into account.
\medskip

For a reaction-rate control system, we can write down the diffusion-reaction equations\,\cite{book-gary} for the concentrations of point vacancies and interstitial atoms as

\begin{eqnarray}
\nonumber
&&\frac{\partial c^j_{v}(z,t)}{\partial t}-D_{v}^j\,\frac{\partial^2 c^j_{v}(z,t)}{\partial z^2}={\cal G}^j_0-\frac{{\cal B}^j_{iv}\,\Omega_j(D^j_{i}+D^j_{v})}
{(a_0^j)^2}\,c^j_{i}(z,t)\,c^{j}_{v}(z,t)\\
\nonumber
&-&\sum\limits_{\ell=4}^{\infty}\frac{{\cal B}^{j+1}_{v}D^{j+1}_{v}a^{j+1}_0}{1-({\cal B}^{j+1}_{v}/2\pi)\ln\{\pi[R^{j+1}_{vd}(\ell)]^2\sigma^{j+1}_{dl}(\ell,t)\}}\,
\sigma^{j+1}_{dl}(\ell,t)\,c^{j+1}_{v}(z,t)\,\delta(z-z_{j+1})\\
&-&\sum\limits_{\ell=4}^{\infty}\frac{{\cal B}^j_{v}D^j_{v}a^j_0}{1-({\cal B}^j_{v}/2\pi)\ln\{\pi[R^j_{vd}(\ell)]^2\sigma^j_{dl}(\ell,t)\}}\,
\sigma^j_{dl}(\ell,t)\,c^j_{v}(z,t)\,\delta(z-z_{j})\ ,
\label{eqn-39}
\end{eqnarray}

\begin{eqnarray}
\nonumber
&&\frac{\partial c^j_{i}(z,t)}{\partial t}-D^j_{i}\,\frac{\partial^2 c^j_{i}(z,t)}{\partial z^2}={\cal G}^j_0-\frac{{\cal B}^j_{iv}\,\Omega_j(D^j_{i}+D^j_{v})}
{(a^j_0)^2}\,c^j_{i}(z,t)\,c^j_{v}(z,t)\\
\nonumber
&-&\sum\limits_{\ell=4}^{\infty}\frac{{\cal B}^{j+1}_{i}D^{j+1}_{i}a^{j+1}_0}{1-({\cal B}^{j+1}_{i}/2\pi)\ln\{\pi[R^{j+1}_{id}(\ell)]^2\sigma^{j+1}_{dl}(\ell,t)\}}\,
\sigma^{j+1}_{dl}(\ell,t)\,c^{j+1}_{i}(z,t)\,\delta(z-z_{j+1})\\
&-&\sum\limits_{\ell=4}^{\infty}\left\{\frac{{\cal B}^j_{i}D^j_{i}a^j_0}{1-({\cal B}^j_{i}/2\pi)\ln\{[\pi[R^j_{id}(\ell)]^2\sigma^j_{dl}(\ell,t)\}}\right\}
\sigma^j_{dl}(\ell,t)\,c^j_{i}(z,t)\,\delta(z-z_{j})\ ,
\label{eqn-40}
\end{eqnarray}
where the small thermal-equilibrium concentration of point vacancies has been neglected at low temperatures,
the terms on the right-hand side of the equations correspond to diffusion sources and reactions,
integer $j$ is the layer index, integer $\ell$ indicates the number of interstitials enclosed within a planar dislocation loop\,\cite{bullough},
$z_{j}$ and $z_{j+1}$ represent the left and right interface positions of the $j$th layer,
$c^j_{v}(z,t)$ and $c^j_{i}(z,t)$ are the concentrations of point vacancies and interstitials,
$D^j_{v}$ and $D^j_{i}$ are the diffusion coefficients, and
${\cal G}^j_0$ is the production rate for Frenkel pairs.
Here, $\rho_d(z)$ can be obtained by multiplying the sample cross-sectional area with $c^j_{v}(z,t)$ and $c^j_{i}(z,t)$.
In addition, in Eqs.\,(\ref{eqn-39}) and (\ref{eqn-40}),
we used the facts that in a reaction-rate control system ${\cal R}^j\equiv\Gamma^j_{i,v}={\cal B}^j_{iv}\,\Omega_jD^j_{i,v}/(a_0^j)^2$ for the vacancy-interstitial recombination rate,
$\Gamma^j_{\{i,v\}d}(\ell,t)=[\kappa^j_{\{i,v\}d}(\ell,t)]^2D^j_{\{i,v\}}/\sigma^j_{dl}(\ell,t)$ is the rate for the interaction between defects and interface dislocation loops,
and $[\kappa^j_{\{i,v\}d}(\ell,t)]^2={\cal B}^j_{\{i,v\}d}(\ell)\,\sigma^j_{dl}(\ell,t)$ for the dislocation loop-sink strength,
where ${\cal B}^j_{iv}$ is the bias factor for recombinations\,\cite{book-gary},
$\Omega_j$ is the atomic volume, $a^j_0$ is the lattice constant, and $c^j_{FP}=\sqrt{{\cal G}^j_0(a_0^j)^2/[{\cal B}^j_{iv}\Omega_j(D^j_{i}+D^j_{v})]}$ is the initial number of Frenkel pairs.
Furthermore, ${\cal B}^j_{\{i,v\}d}(\ell)\sim {\cal B}^j_{i,v}$ and $R^j_{\{i,v\}d}(\ell)\sim \ell a^j_0/2\pi$ are the bias factors for the reactions and the capture radii
of vacancy-dislocation loop ($vd$) and interstitial-dislocation loop ($id$),
${\cal B}^j_{v}\neq {\cal B}^j_{i}$ are the bias factors for vacancies and interstitials, and
finally $\sigma^j_{dl}(\ell,t)$ is the growth-strain-induced interface dislocation-loop (enclosing $\ell$ captured interstitial atoms) areal density.
\medskip

The diffusion coefficients $D^j_{i,v}$ for point vacancies and interstitials can be calculated from\,\cite{book-gary}

\begin{equation}
D^j_{i,v}=\alpha^j\,(a_0^j)^2\,\omega^j_{D}\,\exp\left(-\frac{E_{i,v}^{j}}{k_BT}\right)\ ,
\label{eqn-41}
\end{equation}
where $E_{i,v}^{j}$ are the migration energies for point vacancies and interstitials,
$\alpha^j$ is determined by the diffusion mechanism and crystal symmetry, $\omega^j_{D}=(6\pi^2/\Omega_j)^{1/3}\,v_{s}$ is the Debye frequency and $v_{s}$ is the sound velocity of the host semiconductor.
\medskip

The interface dislocation-loop density $\sigma^j_{dl}(\ell,t)$ in Eqs.\,(\ref{eqn-39}) and (\ref{eqn-40}) can be found from the following reaction equation\,\cite{book-gary} (for $\ell\geq 4$), i.e.

\[
\frac{\partial\sigma^j_{dl}(\ell,t)}{\partial t}=
\left[\beta^j_{v}(\ell+1,t)+\alpha^j_{i}(\ell+1,t)\right]\sigma^j_{dl}(\ell+1,t)+\beta^j_{i}(\ell-1,t)\,\sigma^j_{dl}(\ell-1,t)
\]
\begin{equation}
-\left[\beta^j_{v}(\ell,t)+\beta^j_{i}(\ell,t)+\alpha^j_{i}(\ell,t)\right]\sigma^j_{dl}(\ell,t)\ ,
\label{eqn-43}
\end{equation}
where $\sigma^j_{dl}(\ell,t=0)=\sigma_{0}^j\,\delta_{\ell,4}$ and $\sigma_{0}^j$ is the initial density for the smallest interface dislocation loops containing four interstitials, the absorption [$\beta^j_{i,v}(\ell,t)$]
and the emission [$\alpha^j_{i}(\ell,t)$] rates are given by\,\cite{book-gary}

\begin{equation}
\beta^j_{i,v}(\ell,t)=\ell a^j_0\,{\cal B}^j_{i,v}\,D^j_{i,v}\,c^j_{i,v}(z_j,t)\ ,
\label{eqn-44}
\end{equation}

\begin{equation}
\alpha^j_{i}(\ell,t)=\ell a^j_0\,{\cal B}^j_{i}\left(\frac{D^j_{i}}{\Omega_j}\right)\exp\left[-\frac{E^j_{b,i}(\ell)}{k_BT}\right]\ ,
\label{eqn-45}
\end{equation}
and $E^j_{b,v}(\ell)$ is the binding energy for a planar cluster of $\ell$ interstitials.
\medskip

\begin{figure}
\includegraphics[width=0.75\textwidth]{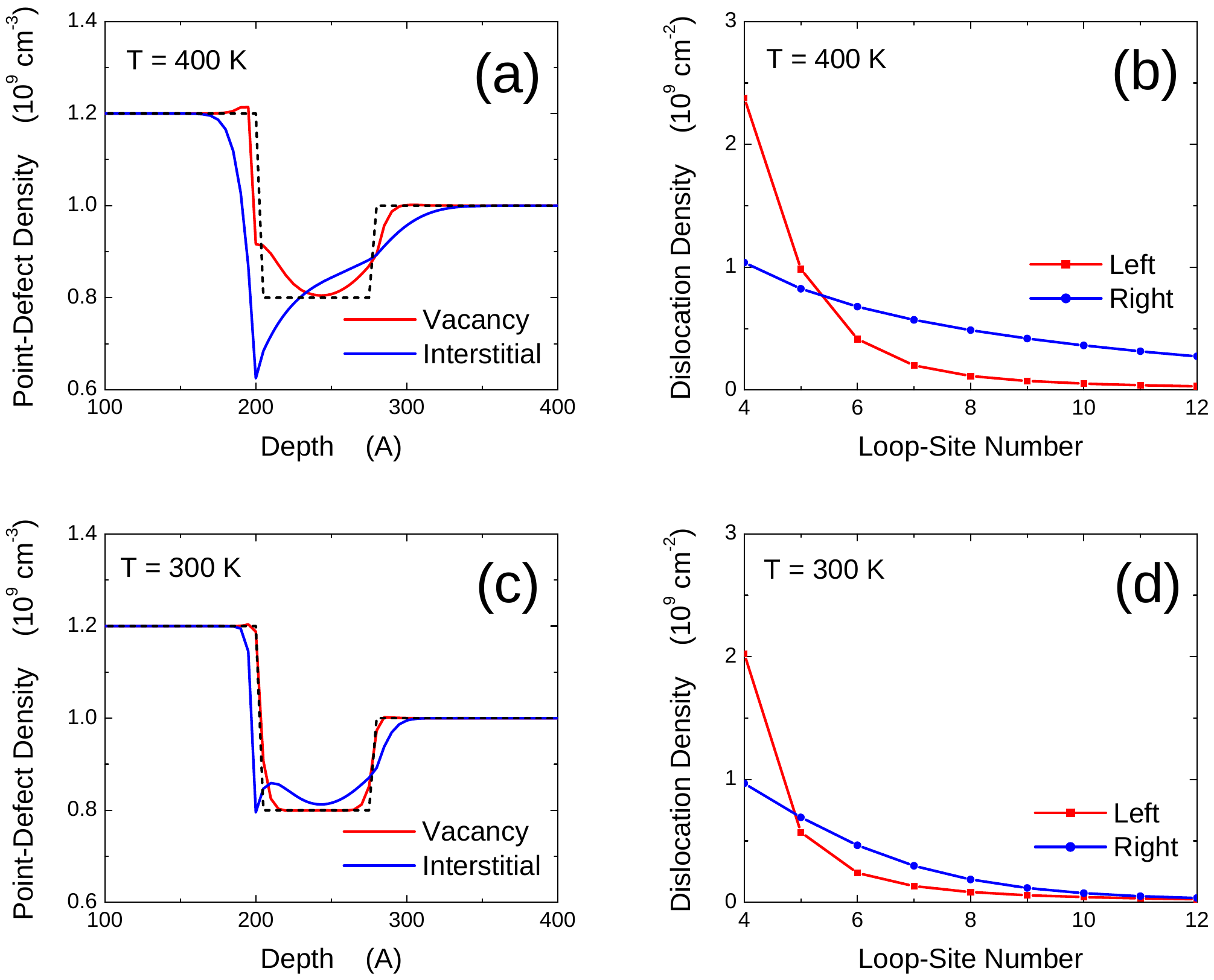}
\caption{(Color online) Concentrations of point vacancies $c^j_v(z)$ and interstitials $c^j_i(z)$ [($a$), ($c$)],
and dislocation-loop densities $\sigma_{dl}^j(\ell)$ at two interfaces [($b$),\,($d$)],
in an AlAs-205\AA/InAs-75\AA/GaAs-255\AA\ single quantum well at $T=400\,$K [($a$),\,($b$)] and $300\,$K [($c$),\,($d$)].
Here, ${\cal G}_0^j$ are $4.6$, $0.9$, $2.1$ in units of $10^{17}\,$cm$^{-3}\,$sec$^{-1}$ and $c^j_{FP}$ are $1.2$, $0.8$, $1.0$ in units of $10^{9}\,$cm$^{-3}$ for $j=1,\,2,\,3$.
In addition, $\sigma_0^j$ are $2.0$, $1.0$ in units of $10^{9}\,$cm$^{-2}$ for $j=1,\,2$. The values for other parameters, i.e., bias factors, absorption and emission rates, diffusion coefficients,
have been obtained from crystal symmetries\,\cite{book-gary} and by scaling melting temperatures with respect to SiC materials\,\cite{gao-1}.}
\label{distri-1}
\end{figure}

We show in Figs.\,\ref{distri-1}($a$) and \ref{distri-1}($c$) the steady-state spatial distributions for concentrations
of point vacancies $c^j_{v}(z)$ and interstitials $c^j_{i}(z)$ in an AlAs/InAs/GaAs single-quantum-well system.
We notice from Eqs.\,(\ref{eqn-39}), (\ref{eqn-40}) and (\ref{eqn-43}) that both  $c^j_{v}(z)$ and $c^j_{i}(z)$ in a steady state eventually become proportional to ${\cal G}^j_0$
although $c^j_{FP}$ is initially proportional to $\sqrt{{\cal G}^j_0}$.
Here, the comparison of results at $T=400\,$K (upper) and $300\,$K (lower) are presented to demonstrate the diffusion of point vacancies into the well through both interfaces due to
thermally-enhanced diffusion coefficients of vacancies. However, the interstitial concentration around the left interface is greatly depleted (deep dip) at $T=400\,$K as a result of
large absorptions by dislocation loops although they still diffuse into the well through the right interface.
In Figs.\,\ref{distri-1}($b$) and \ref{distri-1}($d$), we display results for steady-state distributions of dislocation-loop densities
$\sigma^j_{dl}(\ell)$ ($\ell\geq 4$) as functions of loop-site number, corresponding to the left ($j=1$) and right ($j=2$) interfaces at $T=300$ and $400\,$K. Here,
the increases of dislocation-loop densities ($\ell=4$) at the left interface and the simultaneous swellings of dislocation loops ($\ell>4$) at the right interface are found due to
the enhanced reactions with point interstitials by their increased diffusion coefficients.
Moreover, the defect diffusions occur mainly around interfaces between two adjacent layers or across the interfaces, and $c^j_{v}(z)\neq c^j_{i}(z)$ due to their different diffusion coefficients
although these two concentrations are initially identical.

\subsection{Defect Production by Proton Radiation}
\label{sec-3-2}

The diffusion-reaction equations presented in Sec.\,\ref{sec-3-1} can be applied to find the spatial dependence of the
one-dimensional distribution function $\rho_d(z)$ of defects. However, the initial conditions of these equations
require the production rate and the concentration of proton-produced Frenkel pairs.
Therefore, we must study the production dynamics of point defects under proton irradiation with different kinetic energies,
which connects the lab-measured defect effects ($\propto$ number of point defects) to space-measured energy-dependent proton fluxes in a particular earth orbit.
For this purpose, an atomic-level molecular-dynamics simulation approach is employed  with help from a Tersoff potential fitted by parameters.\,\cite{add10,gao-1}
\medskip

For a bulk material, the production rate per unit volume ${\cal G}_0(E_{i})$ [sec$^{-1}\cdot$cm$^{-3}$] for the displacement atoms in a crystal lattice can be calculated from\,\cite{book-gary}

\begin{equation}
{\cal G}_0(E_{i})=n_{at}\,\sigma_{D}(E_{i})\,{\cal F}_0(E_{i})\ ,
\label{eqn-46}
\end{equation}
where $E_{i}$ [MeV] is the incident proton kinetic energy, $n_{at}$ [cm$^{-3}$] is the crystal atom volume density, ${\cal F}_0(E_{i})$ [cm$^{-2}\cdot$sec$^{-1}$] is the incident energy-dependent proton flux,
and $\sigma_{D}(E_{i})$ [cm$^{2}$] is
the energy-dependent displacement cross section.
\medskip

Physically, the displacement cross section $\sigma_{D}(E_{i})$ in Eq.\,(\ref{eqn-46}) describes the probability for the displacement of struck lattice atoms by incident protons,
therefore, we can directly write

\begin{equation}
\sigma_{D}(E_{i})=\int\limits_{E_{d}}^{\varepsilon_{max}(E_{i})} d\varepsilon_{T}\,{\cal Q}(\varepsilon_{T})\,\sigma_{C}(E_{i},\varepsilon_{T})\,
{\cal N}_{MD}(\varepsilon_{T})\ ,
\label{eqn-47}
\end{equation}
where $\sigma_{C}(E_{i},\varepsilon_{T})$ [cm$^{2}\cdot({\rm keV})^{-1}$] is the differential energy transfer cross section by collision with the lattice, which measures the probability that an incident proton with kinetic energy $E_{i}$ will transfer a recoil energy $\varepsilon_{T}$ [keV] to a struck lattice atom, ${\cal N}_{MD}(\varepsilon_{T})$ [no unit] represents the average number of displaced atoms produced by collision with the lattice,
and $E_{d}$ labels the energy threshold, i.e., the energy required to produce a stable Frenkel pair.
In addition, $\varepsilon_{max}(E_{i})=[4m_0M_0/(m_0+M_0)^2]\,E_{i}$ is the upper limit for the recoil energy gained by the struck lattice atom,
where $M_0$ refers to the mass of lattice atoms and $m_0$ to the mass of incident protons.
\medskip

The function ${\cal Q}(\varepsilon_{T})$ [unitless] introduced in Eq.\,(\ref{eqn-47}) is the so-called Lindhard partition function and is written as\,\cite{niel,niel1,niel2}

\begin{equation}
{\cal Q}(\varepsilon_{T})=\frac{1}{1+{\cal K}_{L}\,g(\varepsilon_{T}/E_{L})}\ ,
\label{eqn-48}
\end{equation}
where the Ziegler-Biersack-Littmark (ZBL) reduced-energy $E_{L}$ is defined as

\begin{equation}
E_{L}=\left(\frac{m_0+M_0}{M_0}\right)\frac{Z_1Z_2e^2}{4\pi\epsilon_0a_{u}}\ ,
\label{eqn-49}
\end{equation}
while the reduced electronic energy-loss factor ${\cal K}_{L}$ is

\begin{equation}
{\cal K}_{L}=\frac{Z_1^{2/3}Z_2^{1/2}}{12.6\left(Z_1^{2/3}+Z_2^{2/3}\right)^{3/4}}\,\frac{[(1+(M_0/m_0)]^{3/2}}{\sqrt{M_0/m_0}}\ ,
\label{eqn-50}
\end{equation}
$m_0=1.67\times 10^{-27}\,$kg is the proton mass,
$a_{u}=0.8853\,a_{B}/\left(Z_1^{2/3}+Z_2^{2/3}\right)^{1/2}$ is the ZBL universal screening length, $a_{B}=4\pi\epsilon_0\hbar^2/m_{e}e^2=0.5292\,$\AA\, is the Bohr radius,
$m_{e}$ is the free-electron mass, and the Lindhard function $g(x)$ is calculated as

\begin{equation}
g(x)=x+0.40244\,x^{3/4}+3.4008\,x^{1/6}\ .
\label{eqn-51}
\end{equation}
In the current case, we set $Z_1=1$ (proton), $Z_2=31$ (Ga) or $33$ (As) for the nuclear charge number of lattice atoms.
\medskip

Moreover, the differential energy transfer cross section $\sigma_{C}(E_{i},\varepsilon_{T})$ [cm$^{2}\cdot({\rm keV})^{-1}$] can be approximated as\,\cite{niel2}

\begin{equation}
\sigma_{C}(E_{i},\varepsilon_{T})=-\frac{\pi a_{u}^2}{2}\,\alpha_{s}^2(E_{i})\,\frac{h_0([\tau(E_{i},\varepsilon_{T})]^{1/2})}{[\tau(E_{i},\varepsilon_{T})]^{3/2}\,\varepsilon_{max}(E_{i})}\ ,
\label{eqn-52}
\end{equation}
where $\tau(E_{i},\varepsilon_{T})=\alpha_{s}^2(E_{i})\varepsilon_{T}/\varepsilon_{max}(E_{i})$ is the dimensionless collision parameter, $\alpha_{s}(E_{i})=E_{i}/E_{L}$
is the scaled ZBL reduced energy.
The function $h_0(x)$ introduced in Eq.\,(\ref{eqn-52}) is defined as

\begin{equation}
h_0(x)=\frac{\ln A(x)}{2B(x)}+\frac{ax}{2A(x)B(x)}-\frac{x\ln A(x)\left(1+bc\,x^{c-1}+d/2x^{1/2}\right)}{2B^2(x)}\ ,
\label{eqn-53}
\end{equation}
where $A(x)=1+ax$, $B(x)=x+bx^c+dx^{1/2}$, $a=1.1383$, $b=0.01321$, $c=0.21226$ and $d=0.19593$ are four parameters.
\medskip

\begin{figure}
\includegraphics[width=0.45\textwidth]{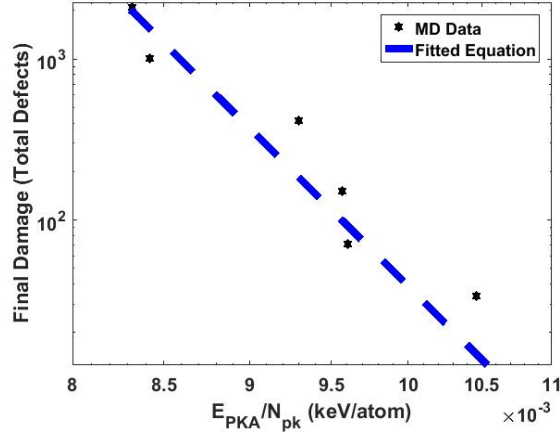}
\caption{(Color online) Total number ${\cal N}_{MD}(\varepsilon_{T})$ of defects in MD simulation as a function of the PKA energy $E_{PKA}\equiv\varepsilon_{T}$ scaled by the number $N_{pk}$
of defects at the peak time in Fig.\,\ref{gao1},
where the formula ${\cal N}_{MD}(\varepsilon_{T})={\cal A}_0[\varepsilon_{T}({\rm keV})]^n$ and $N_{pk}=B_0[\varepsilon_{T}({\rm keV})]^\ell$ with parameters ${\cal A}_0=75.0722134597135$, $n=1.11078028052446$, $B_0=64.1418065233329$, and $\ell=1.0391196703896$ extracted from fitting (dashed curve).}
\label{gao3}
\end{figure}

Finally, ${\cal N}_{MD}(\varepsilon_{T})$ in Eq.\,(\ref{eqn-47}) can be computed by using MD simulations. As shown in Fig.\,\ref{gao3}, the calculated ${\cal N}_{MD}(\varepsilon_{T})$
can be fitted reasonably well by a simple power law,
i.e., ${\cal N}_{MD}(\varepsilon_{T})={\cal A}_0[\varepsilon_{T}({\rm keV})]^n$ with proper fitting parameters ${\cal A}_0$ and $n$.
Finally, by combining together the results in Eqs.\,(\ref{eqn-46})-(\ref{eqn-52}), for a given flux spectrum ${\cal F}_0(E_{i})$ we get the production rate ${\cal G}_0(E_i)$ per unit volume as

\[
{\cal G}_0(E_{i})=-\frac{n_{at}\pi a_{u}^2{\cal A}_0}{2}\left[\frac{\alpha_{s}^2(E_{i}){\cal F}_0(E_{i})}{\,\varepsilon_{max}(E_{i})}\right]
\]
\begin{equation}
\times\int\limits_{E_{d}}^{\varepsilon_{max}(E_{i})} d\varepsilon_{T}\,{\cal Q}(\varepsilon_{T})\,[\varepsilon_{T}({\rm keV})]^n\,
\frac{h_0([\tau(E_{i},\varepsilon_{T})]^{1/2})}{[\tau(E_{i},\varepsilon_{T})]^{3/2}}\ ,
\label{eqn-54}
\end{equation}
which can be evaluated numerically once fitting parameters ${\cal A}_0$ and $n$ are obtained.
Here, ${\cal G}_0(E_{i})$ is related to the more familiar non-ionizing energy loss\,\cite{add-4} ${\rm NIEL}(E_{i})$ by  ${\cal G}_0(E_{i})=(\rho_{at}/n_{at})\,(0.4/E_{d})\,{\cal F}_0(E_{i})\,{\rm NIEL}(E_{i})$
with $\rho_{at}$ being a crystal atom weight density.\,\cite{add-4} Furthermore, the concentration $c_{FP}(E_i)$ for Frenkel-pair defects can be roughly estimated from
$c_{FP}(E_i)={\cal G}_0(E_{i})\,(\tau_0+\tau_t/2)$, where $\tau_t$ is the effective proton transit time through the sample, and $\tau_0\sim 10\,$ns, which is proportional to $1/\sqrt{{\cal F}_0(E_{i})}$,
is the time required to reach a steady state for generation of Frenkel-pair defects after the production has been balanced by the recombination.
\medskip

\begin{figure}
\includegraphics[width=0.45\textwidth]{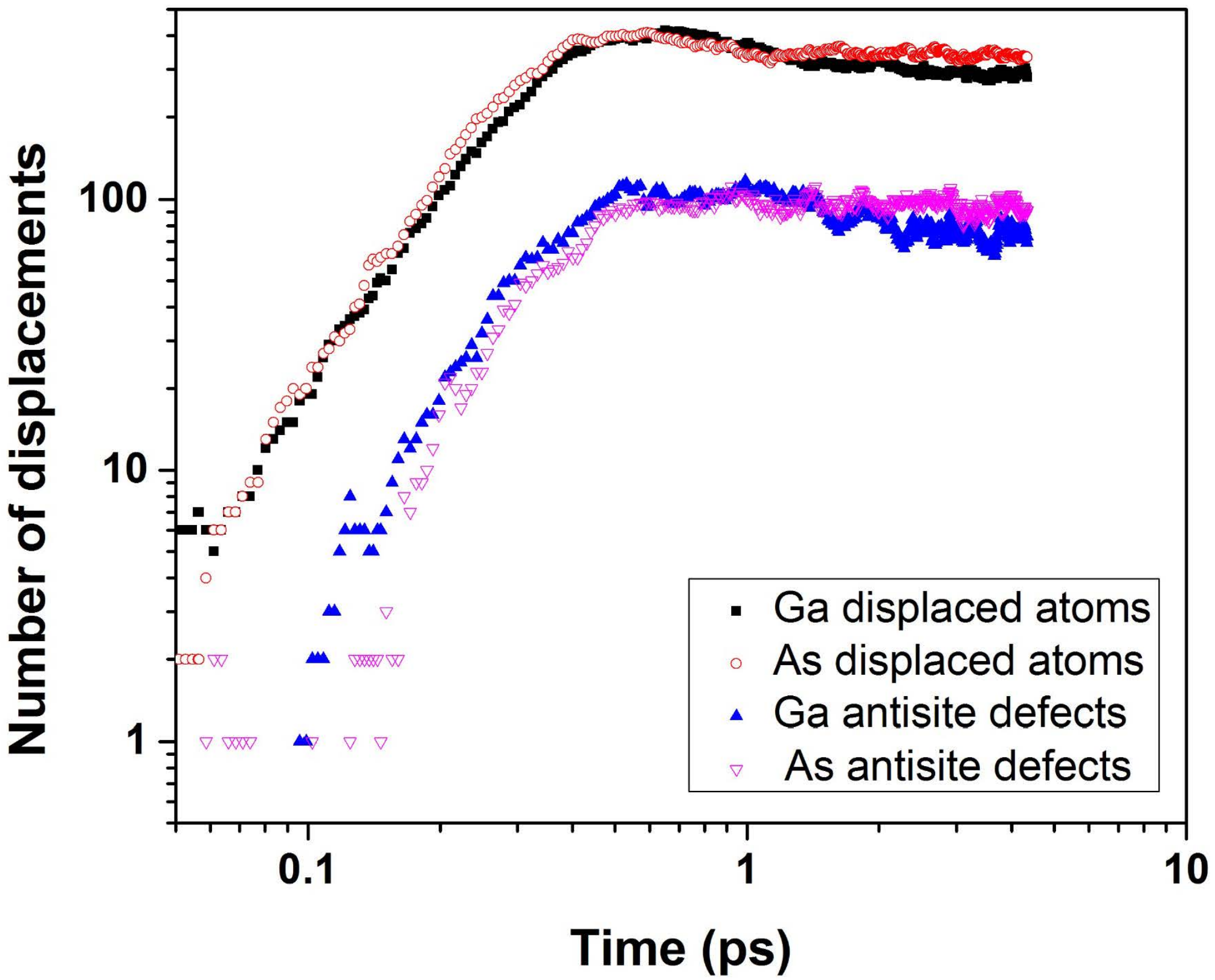}
\caption{(Color online) The number of Ga and As displaced atoms and antisite defects as a function of time in a $\varepsilon_{T}=10\,$keV Ga-PKA cascade in GaAs, where the peak time about $0.8\,$ps is found.}
\label{gao1}
\end{figure}

We present in Fig.\,\ref{gao1} the numerical results for calculated number of lattice-atom displacements as a function of time after a Ga PKA has been introduced to a GaAs
crystal with the recoil energy $\varepsilon_{T}=10\,$keV. From Fig.\,\ref{gao1}, we find that the number of lattice-atom displacements reaches a peak value $N_{pk}$ at about $t=0.8\,$ps.
After this peak time, only $13\%$ of the displaced atoms recombine with vacancies, and most anti-site defects are generated during the collisional phase. In addition, a steady state with
$\varepsilon_{T}=10\,$keV has been reached for $t>10\,$ps,
where As defects are slightly higher than that of Ga defects due to the smaller formation energy for As defects\,\cite{add-5}.
\medskip

\begin{figure}
\includegraphics[width=0.45\textwidth]{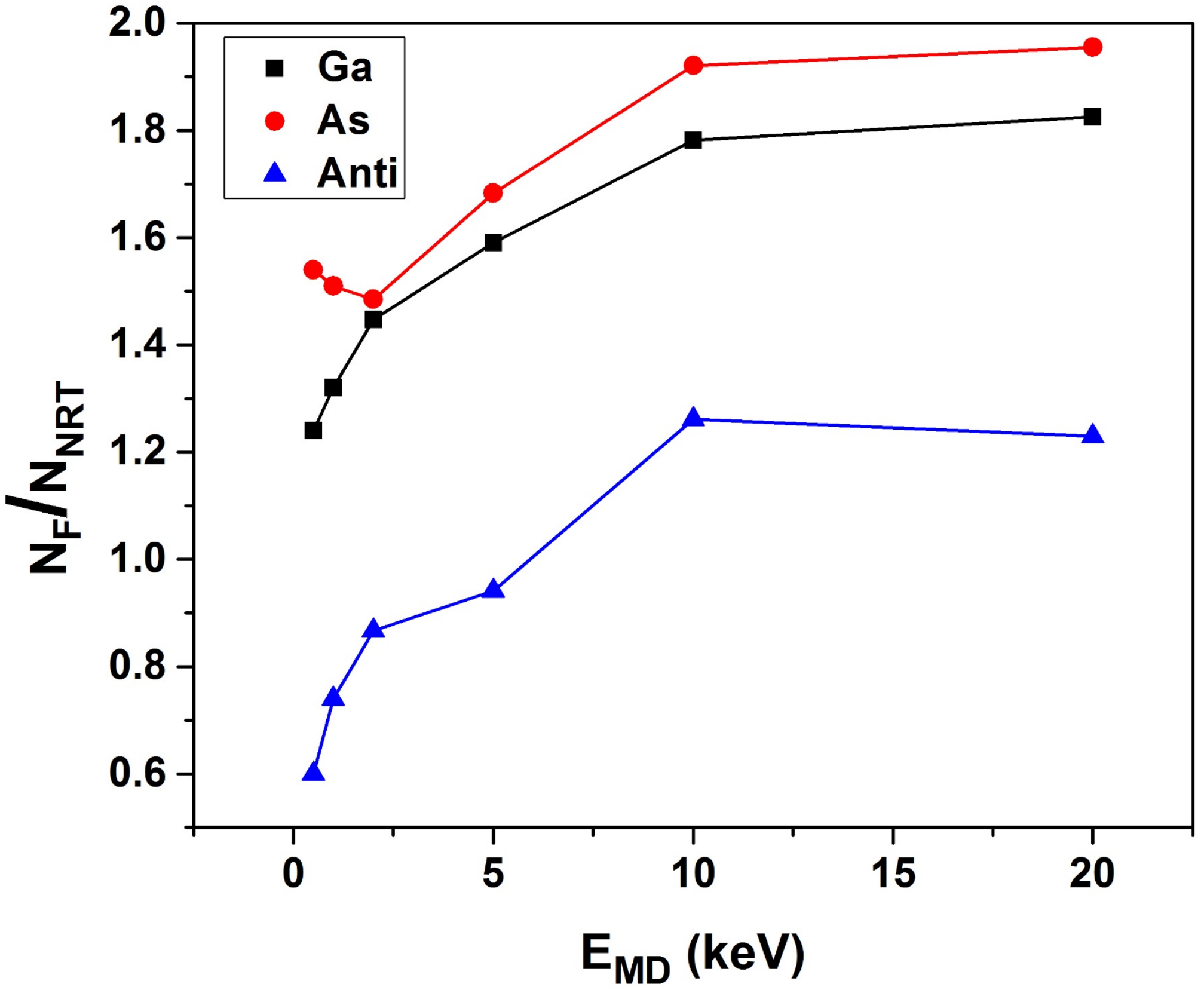}
\caption{(Color online) The number ${\cal N}_{F}(\varepsilon_{T})$ of Ga and As displaced atoms and antisite defects as a function of recoil energy $E_{MD}\equiv\varepsilon_{T}$ in MD simulation at $t=10\,$ps, where the NRT result is given by
${\cal N}_{F}(\varepsilon_{T})=0.8\,\varepsilon_{T}/2E_{d}$.}
\label{gao2}
\end{figure}

The numerical results for the number ${\cal N}_{F}(\varepsilon_{T})$ of Ga and As displaced atoms and anti-site defects as a function of recoil energy $\varepsilon_{T}$ at $t=10\,$ps are displayed in Fig.\,\ref{gao2},
where the NRT result is given by ${\cal N}_{F}(\varepsilon_{T})=N_{NRT}(\varepsilon_{T})\equiv 0.8\,\varepsilon_{T}/2E_{d}$. It is clear from this figure that the number of defects in steady state is found to be
much high than that given by the NRT value. Moreover, nonlinear dependence on $\varepsilon_{T}$ is limited only for low-energy PKA recoils.
\medskip

\begin{figure}
\includegraphics[width=0.45\textwidth]{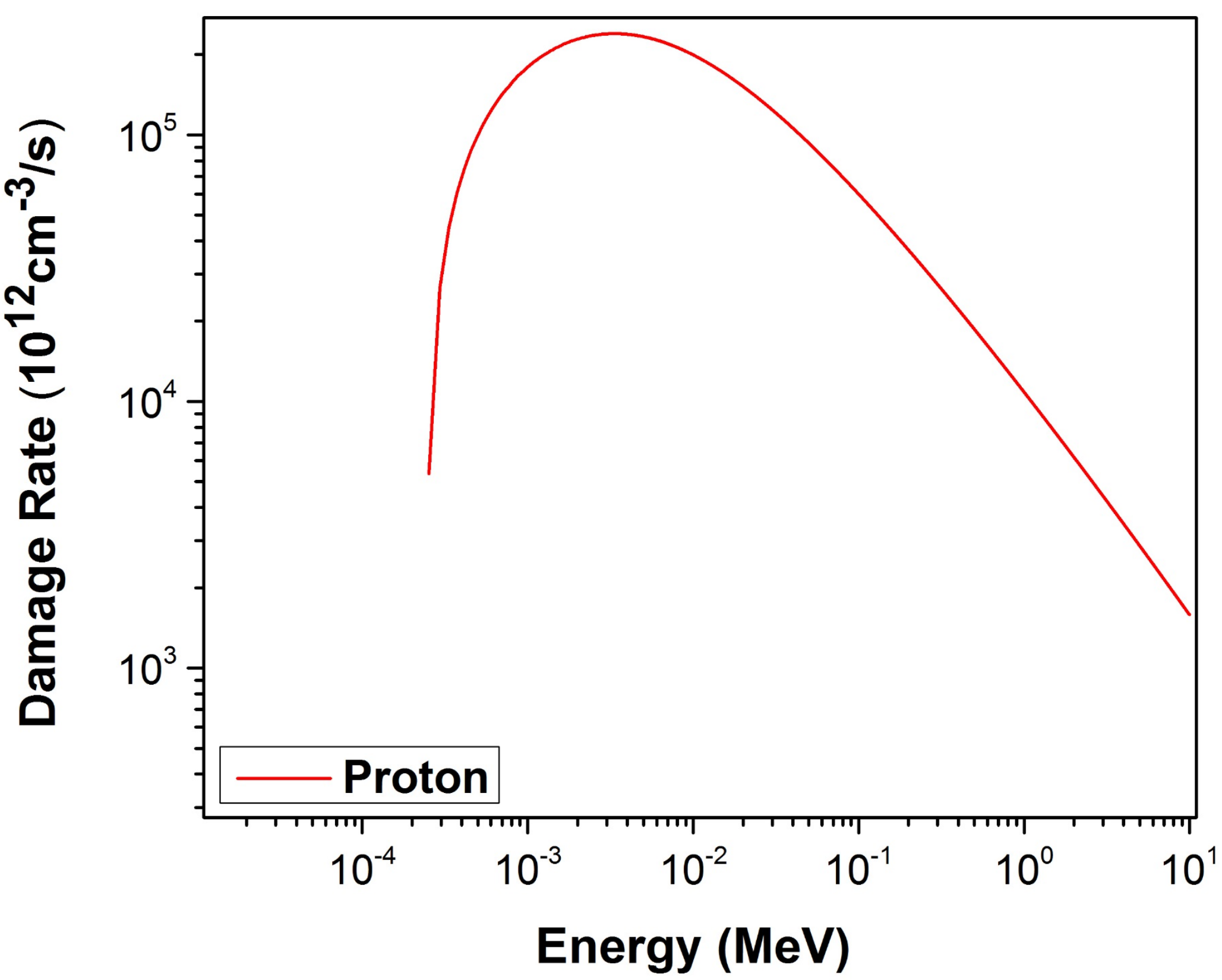}
\caption{(Color online) Calculated defect production (or damage) rate ${\cal G}_0(E_{i})$ per unit volume as a function of the incident-proton kinetic energy $E_{i}$.
Here, the proton flux is assumed to be a constant $3.0\times 10^{12}\,$cm$^{-2}\cdot$sec$^{-1}$.}
\label{gao4}
\end{figure}

In order to provide initial Frenkel-pair defect concentrations and its production rate, we show in Fig.\,\ref{gao4} the numerical result of Eq.\,(\ref{eqn-54}) for ${\cal G}_0(E_{i})$.
It is clear from this figure that there exists a peak for ${\cal G}_0(E_{i})$ as a function of incident proton kinetic energy due to competition between increasing $\varepsilon_{max}(E_i)$
and decreasing $\sigma_C(E_i,\varepsilon_T)$ at the same time.

\section{Summary and Remarks}
\label{sec-4}

In conclusion, we have investigated the effects of point defects on the loss of either electron kinetic energy or incident photons in a multi-quantum-well system.
The influence of proton-radiation-produced defects is taken into account by
applying the vertex correction to a bare polarization function of electrons in quantum wells within the ladder approximation, which goes beyond the usual second-order Born approximation.
Both intralayer and interlayer dynamical screenings to the defect-electron interaction have also been considered under the random-phase approximation.
Furthermore, the defect effects on the electron-energy loss function, as well as on intrasubband and intersubband optical absorption, have been shown and discussed.
\medskip

To find the distribution function of point defects in a layered structure for calculations of defect effects,
we have applied the diffusion-reaction-equation method, where the reactions of point defects with the growth-induced dislocation loops on interfaces of the multi-layered system have been included,
and the increase and decrease of dislocation-loop density and point-defect concentrations were found at the same time
due to thermal enhancement of defect diffusion. In addition, the Frenkel-pair defect production rate and the initial concentration of Frenkel pairs were obtained from
an atomic-level molecular-dynamics model after fitting the numerical results for Frenkel pairs as a function of energy of a primary knock-on atom.
\medskip

For the first time, the defect effect, diffusion-reaction and molecular dynamics models presented in this paper
can be combined with a space-weather forecast model\,\cite{swf,cooke} which predicts spatial-temporal fluxes and particle velocity distributions.
With this combination of theories, the predicted irradiation conditions for particular satellite orbits allow electronic and optoelectronic devices to be specifically designed for operation in space with
radiation-hardening considerations\,\cite{ajss} (such as self-healing and mitigation). This approach will effectively extend the lifetime of satellite onboard electronic and optoelectronic devices
in non-benign orbits and greatly reduce the cost.

\begin{acknowledgements}
This material is based upon work supported by the Air Force Office of Scientific Research (AFOSR).
We would also like to thank Dr. Paul D. LeVan and Dr. Sanjay Krishna for helpful discussions.
\end{acknowledgements}

\newpage

\end{document}